\documentclass[preprint,showpacs,preprintnumbers,amsmath,amssymb]{revtex4}

\usepackage{graphicx}
\usepackage{dcolumn}
\usepackage{bm}

\begin{document}

\newcommand{\balpha}{\bm{\alpha}}
\newcommand{\bmu}{\bm{\mu}}
\newcommand{\bsig}{\bm{\sigma}}
\newcommand{\brho}{\bm{\rho}}
\newcommand{\blam}{\bm{\lambda}}
\newcommand{\bL}{\bm{L}}
\newcommand{\bA}{\bm{A}}
\newcommand{\bB}{\bm{B}}
\newcommand{\bC}{\bm{C}}
\newcommand{\bS}{\bm{S}}
\newcommand{\bx}{\bm{x}}
\newcommand{\by}{\bm{y}}
\newcommand{\bz}{\bm{z}}
\newcommand{\mb}{\textbf m}
\newcommand{\pb}{\textbf p}
\newcommand{\rb}{\textbf r}
\newcommand{\sbd}{\textbf s}

\title{Analysis of dynamical corrections to baryon magnetic moments}

\author{Phuoc Ha}%
\email{pdha@creighton.edu}
\affiliation{%
Department of Physics, Creighton University \\
             Omaha, Nebraska 68178, USA
}%

\author{Loyal Durand}%
 \email{ldurand@theory1.physics.wisc.edu}
\affiliation{%
Department of Physics, University of Wisconsin-Madison \\
             Madison, Wisconsin 53706, USA
}%

\date{\today}

\begin{abstract}
We present and analyze QCD corrections to the baryon magnetic
moments in terms of the one-, two-, and three-body operators which
appear in the effective field theory developed in our recent
papers. The main corrections are extended Thomas-type corrections
associated with the confining interactions in the baryon. We
investigate the contributions of low-lying angular excitations to
the moments quantitatively and show that they are completely
negligible. When the QCD corrections are combined with the
non-quark model contributions of the meson loops, we obtain a
model which describes the moments within a mean deviation of 0.04
$\mu_N$. The nontrivial interplay of the two types of corrections to
the quark-model moments is analyzed in detail, and explains why
the quark model is so successful. In the course of these
calculations,  we parametrize the general spin structure of the
$j=\frac{1}{2}^+$ baryon wave functions in a form which clearly
displays the symmetry properties and the internal angular momentum
content of the wave functions, and allows us to use spin-trace
methods to calculate the many spin matrix elements which appear in
the expressions for the moments. This representation may be useful
elsewhere.
\end{abstract}

\pacs{PACS Nos: 13.40.Em,11.30.Rd}

\maketitle

\section{Introduction}

Baryon magnetic moments ( baryon moments for short)
have been studied intensively for many years using different models and
approaches \cite{M1,RSS,F,RGG,Caldi,GSS,M2,K,J,L,M3,D,Mei,DH-ChPTmoments,Durand,%
DH-loop-moments,M4,DHJ1,DHJ2}. It initially appears somewhat surprising that the
simple, nonrelativistic quark model (QM) \cite{M1,RSS,F,RGG} is so good, with
more complicated models giving only small improvements in the quantitative
description of the baryon moments. This is now understood.

In our recent work on the connections between the QM and effective field
theory \cite{DHJ1,DHJ2}, we showed that the constituent QM for the baryon
moments can be regarded as a rewriting of the relativistic chiral effective
field theory (EFT) in which only the one-body moment operators are retained,
a result also obtained by Morpurgo using his general parametrization method
\cite{M2,M3,M4} derived exactly from QCD. Like Morpurgo, we found that the
success of the additive QM is due to the numerical dominance of the one-body
operators over the nonadditive two- and three-body operators, and suggested
why the latter operators should, in fact, be small.

In our earlier dynamical model \cite{Durand}, an initial baryon
moment operator was  derived from quenched QCD using the Wilson-loop approach
of Brambilla {\em et al.} \cite{brambilla}. This appears as a sum of single-particle
moments $\mu_i\approx\left<eQ_i/2E_i\right>$, where $E_i$ is
the kinetic energy of quark $i$, and a set of Thomas precession terms connected to
the binding interactions. In \cite{DH-loop-moments},
we studied the further corrections to the baryon moments from
from meson loops, which are absent in quenched QCD. This combined approach
substantially improved the agreement
between the theory and experiment, leaving an average difference between the
theoretical and experimental octet moments
of only 0.05$\mu_N$. We did not, however, analyze the components of the model,
or its description in EFT, in detail. The central idea in the context of our
later work on EFT \cite{DHJ1,DHJ2} was to use a dynamical model, namely the
QCD-based quark model, to obtain reasonable estimates of the input parameters
connected to the one- and higher- body operators in the effective field theory,
then to add the meson loop corrections using chiral perturbation theory.

Our goals here are (i) to present the general structure of the baryon moments in
effective field theory, parametrized in a form which connects easily to dynamical
ideas; (ii) to analyze the contributions of meson loops and the dynamical QCD
corrections in terms of the EFT; (iii) sketch the relevant parts
of our dynamical calculations, including our analysis of the contributions of
low-lying angular excitations; and (iv) to show how the loop and dynamical
corrections combine to give a good description of the moments.
Items (ii) and (iii) involve the structure
of the wave functions used to estimate the dynamical corrections, so we give
their form and the methods of calculation
used. Some of this material is potentially useful in other contexts.

The paper is organized as follows: In Sec. \ref{subsec:chiral_moments}, we
briefly review the structure of the baryon moments in the effective field
theory, where the QM moments arise from one-body operators. In
Sec.\ \ref{subsec:background} we sketch a
derivation of the QM moments and further two- and three-body dynamical
corrections from QCD using a Wilson-loop
approach \cite{Durand,DH-loop-moments}. We determine the structure of
the $j=\frac{1}{2}^+$ baryon wave functions needed to calculate these
corrections and develop useful the trace methods for the calculation
in Sec. \ref{sec:wavefuncs}. We analyze the contributions of meson loops,
the dynamical QCD corrections, and internal orbital
angular momenta numerically in Sec. \ref{sec:calculation}, and show why
both the loop and dynamical corrections are needed to obtain a good
description of the data. We present our conclusions and discussion in
Sec. \ref{sec:conclusions}, and give some additional information, such as
proofs and sample calculations, in the appendices.


\section{\label{sec:operators}Theory of the moments}

\subsection{\label{subsec:chiral_moments}Moments in chiral perturbation theory}

In two earlier papers \cite{DHJ1,DHJ2}, we analyzed the structure of the baryon
magnetic moment operators to O($m_s$) in heavy-baryon chiral perturbations
theory (HBChPT), connecting the general spin-flavor structure to the spin
dependence of the underlying dynamical theory. The analysis was done using
flavor-index labeling of the effective baryon fields $B_{ijk}^\gamma(x)$,
where $i,j,k\in u,d,s$, and $\gamma$ is a Dirac spinor index. The transformation
properties of the fields are the same as those of the operators
\begin{equation}
\label{Bijk}
B_{ijk}^\gamma \cong \frac{1}{6}\epsilon_{abc}q_i^{\alpha a}q_j^{\beta b}
q_k^{\gamma c}(C\gamma^5)_{\alpha\beta}
\end{equation}
constructed in terms of free quark operators which carry the color, flavor,
and spin structure of the baryon. This expression for the relativistic fields
reduces in the rest frame of the heavy baryon to the nonrelativistic structure
familiar in the nonrelativistic quark model,
\begin{equation}
\label{Bijk_rest}
B_{ijk}^\gamma\rightarrow \frac{1}{6}\epsilon_{abc}
\left(q_i^{a\,T}i\sigma_2q_j^b\right)q_k^{\gamma c},
\end{equation}
and provides a straightforward connection to semirelativistic dynamical models for
the baryons. It also connects directly to Morpurgo's general parametrization method
for specifying the most general spin and flavor structure of matrix elements in
QCD \cite{M2}. In the following, we will refer to the structure in terms
of ``quarks'' for convenience, but emphasize that the description is completely
equivalent to a relativistic effective field theory.

\begin{figure}
\includegraphics{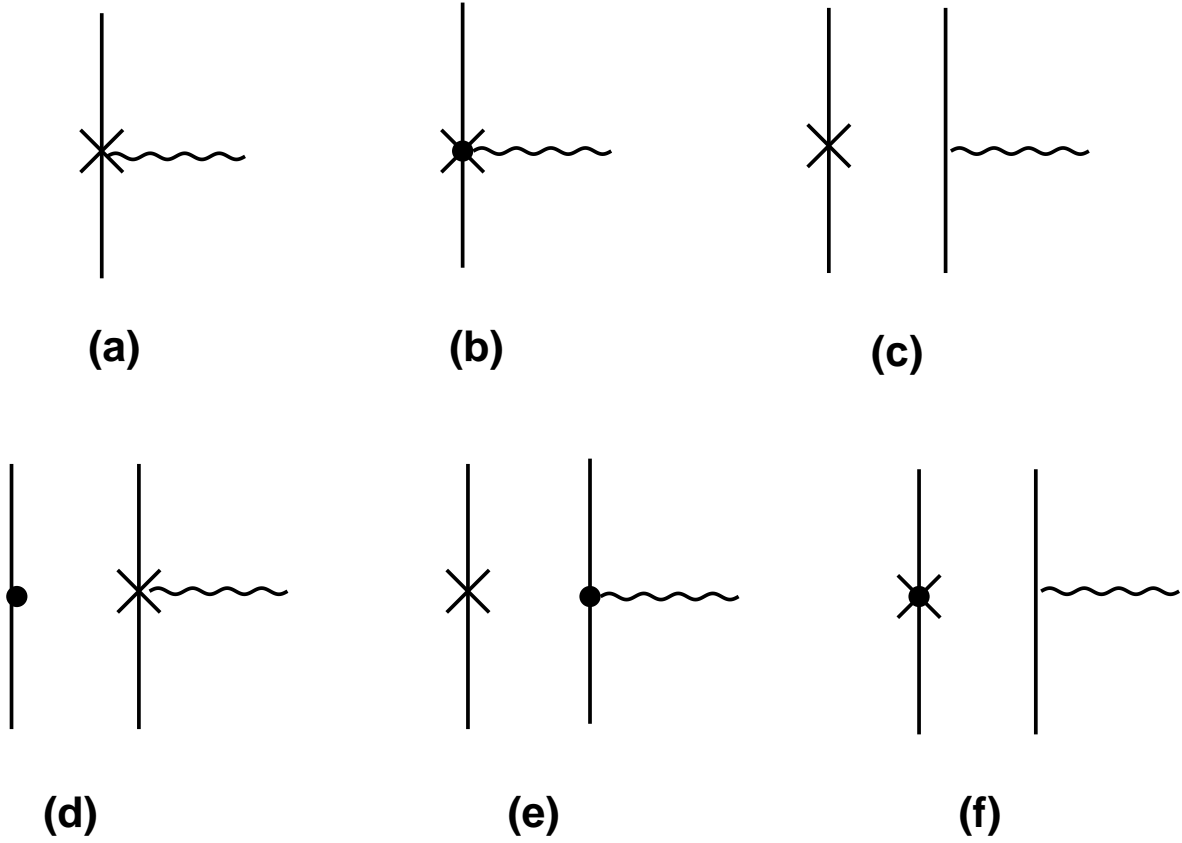}
\caption{\label{fig1} One- and two-body contributions to the baryon magnetic moments
at the quark level. Lines with small wiggles represent photons. There is a factor
$\bsig\!\cdot\!{\textbf B} \leftrightarrow -\sigma^{\mu\nu} F_{\mu\nu}$
in the Lagrangian at the quark-photon vertex, where $\textbf B$ is the external magnetic
field. Crosses correspond to factors of the quark charge matrix $Q$, and dots to
insertions of the flavor matrix $M$. Graphs (a) and (b) give the SU(6) structure for
the moments.}
\end{figure}

\begin{figure}
\includegraphics{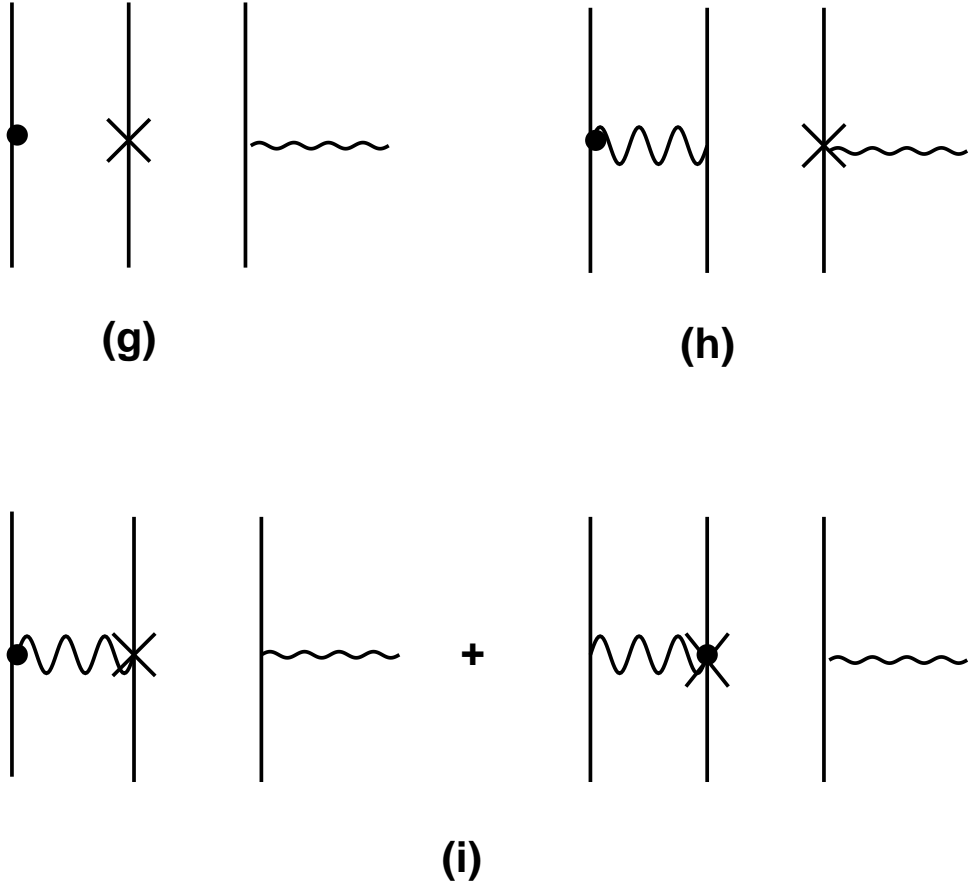}
\caption{\label{fig2} Three-body contributions to the baryon magnetic moments. The
two graphs in (i) appear together with the same overall coefficient. The structures in
graphs (h) and (i) are reducible to those in Figs.\ \protect\ref{fig1} and
\protect\ref{fig2}\,(g).}
\end{figure}

As shown in \cite{DHJ2}, the octet moments $\bmu$ can be
written through first order in the chiral symmetry breaking parameter $m_s$ in
terms of seven operators ${\mb}_r$ which correspond
to the the diagrams (a)-(g) in Figs.\ \ref{fig1} and \ref{fig2}. The contribution
of the octet moments to the chiral Lagrangian for is given to O($m_s$) as a matrix
element of the form $(\bar{B}\bmu B)\cdot{\bf B}$ where
\begin{equation}
\label{mu_r}
\bmu = \sum_r \mu_r {\mb}_r
\end{equation}
is a combination of the  quark-level operators
\begin{eqnarray}
{\mb}_a &=& \sum_i\, (Q\bsig)_i, \\
{\mb}_b &=& \sum_i\, (QM\bsig)_i, \\
{\mb}_c &=& \sum_{i\not=j}\,Q_i\bsig_j, \\
\label{quark-level_moments}
{\mb}_d &=& \sum_{i\not=j}\,M_i(Q\bsig)_j, \\
{\mb}_e &=& \sum_{i\not=j}\,Q_i(M\bsig)_j, \\
{\mb}_f &=& \sum_{i\not=j}\,(QM)_i\bsig_j, \\
{\mb}_g &=& \sum_{i\not=j\not=k}\,M_iQ_j\bsig_k.
\end{eqnarray}
$Q$ is the diagonal quark charge matrix, $Q=\textrm{diag} (2/3,\,-1/3,\,-1/3)$,
and $M$ is the diagonal mass matrix for the strange quark, $M=\textrm{diag}(0,\,0,\,1)$.
A flavor index attached to a $Q$ or $M$ means that the matrices are taken to act on
that flavor quark so that, for example, $\bar{B}(Q\bsig)_iB$ is the matrix element
$\bar{B}_{k'j'i'}Q_{i'i}\delta_{j'j}\delta_{k'k}\bsig_iB_{ijk}$, with $\bsig_i$
acting on quark $i$ in the representation in Eq.\ (\ref{Bijk_rest}),
$q_i\rightarrow\bsig_iq_i$.
These operators are shown diagrammatically in Figs. \ref{fig1} and \ref{fig2}
(Figures 10 and 11 in \cite{DHJ2}). The expressions for the baryon moments in terms
of the $\mu_r$ is given in Table \ref{table:B-mu_r} \footnote{The connection of
the coefficients $\mu_b$ and $\mu_d$ to the coefficients $\mu_1$-$\mu_7$ of the
alternative operators defined in Eqs. (4.6)-(4.12) of \cite{DHJ2}, are unfortunately
given incorrectly in Eq.\ (4.14) in that paper. The correct relations are
$\mu_b=\mu_2+\mu_4+\mu_5+\mu_6+\mu_7$ and $\mu_d=\mu_4+\mu_7$. The expressions
for the baryon moments in terms of $\mu_1$-$\mu_7$ are simpler than those given
here in Table \ref{table:B-mu_r}, but the simple connection with the underlying
dynamics is lost.}.

\begin{table}
\caption{The coefficients for the expression of the octet baryon moments in terms
of the operators corresponding to the diagrams in Figs.\ \ref{fig1} and \ref{fig2},
labelled as in the figures, plus the operator
${\mb}_{MM} = \sum_{i\not=j}[Q_iM_jM_k-(QM)_iM_j]$ of second order in the
strange-quark mass matrix $M$. The moment of the $\Sigma^0$ is determined by
isospin invariance, $\mu_{\Sigma^0}=\frac{1}{2}(\mu_{\Sigma^+}+\mu_{\Sigma^-})$.}
\label{table:B-mu_r}
\begin{ruledtabular}
\begin{tabular}{lcccccccc}
Baryon & $\mu_a$ & $\mu_b$ & $\mu_c$ & $\mu_d$ & $\mu_e$ & $\mu_f$ & $\mu_g$ & $\mu_{MM}$ \\
\hline
$p$ & 1 & 0 & 0 & 0 & 0 & 0 & 0 & 0 \\
$n$ & -2/3 & 0 & 2/3 & 0 & 0 & 0 & 0 & 0 \\
$\Sigma^+$ & 1 & 1/9 & 0 & 8/9 & -4/9 & -4/9 & 8/9 & 0 \\
$\Sigma^-$ & -1/3 & 1/9 & -2/3 & -4/9 & 2/9 & -4/9 &-4/9 & 0 \\
$\Xi^0$ &-2/3 & -4/9 & 2/3 & -8/9 & 4/9 & -2/9 &10/9 & 1 \\
$\Xi^-$ & -1/3 & -4/9 & -2/3 & -2/9 & -8/9 & -2/9 & -2/9 & 0 \\
$\Lambda$ & -1/3 & -1/3 & 1/3 & 0 & 1/3 & 0 & 0 & 0\\
$\Sigma^0\Lambda$ & $1/\sqrt{3}$ & 0 & $-1/\sqrt{3}$ & $1/\sqrt{3}$ & 0
& 0 & $-1/\sqrt{3}$ & 0
\end{tabular}
\end{ruledtabular}
\end{table}

The coefficients $\mu_r$ of the operators above are not specified in HBChPT, and
the effective field theory is not predictive without further input. The additive
quark model for the moments involves only the one-body operators ${\mb}_a$ and
${\mb}_b$, Figs.\ \ref{fig1}\,(a) and (b). These combine to give the effective
QM moment operator $\mu_aQ+\mu_bQM$. This describes the moments quite well, with a
root-mean-square deviation of theory from experiment of 0.12 $\mu_N$, about
11\% of the average magnitudes of the moments. The underlying reasons for this
striking success have been discussed in \cite{DHJ1,DHJ2} from the point of view
of the weak spin dependence of the interactions in dynamical models, and in
\cite{M2,M3,M4} in a QCD-based analysis. These related analyses conclude that
the two-body contributions to the moments involving ${\mb}_c$-${\mb}_f$ and
the three-body contributions proportional to ${\mb}_g-{\mb}_i$ should all be
small.  Thus, graph \ref{fig1}\,(c) gives a Thomas-type
contribution to the moments from the spin dependence of the quark interactions as
discussed in \cite{DH-loop-moments} and \cite{DHJ1}. Graph \ref{fig1}\,(d) describes
the effect at O($m_s$) of the strange-quark mass on the
leading contribution to the moment from a different quark, graph \ref{fig1}\,(a),
through the mass dependence of the wave function, while \ref{fig1}\,(e) and
\ref{fig1}\,(f) give the corresponding effects on the Thomas term. Graph (g) in
Fig.\ \ref{fig2} gives the three-body strange-mass correction to the Thomas term.
The extra three-body operators (h) and (i) arise from the effect of spin-spin
interactions on the leading and Thomas terms at O($m_s$). These appear in a combined
treatment of the octet and decuplet moments, but are reducible to the preceding terms
if we consider only the octet.

The dynamical problem  in understanding the baryon moments is the actual calculation
of the coefficients $\mu_r$, and of possible terms of O($m_s^2$) and O($m_s^3$) which
could affect the $\Xi$, $\Xi^*$, and $\Omega^-$ moments. The origin of the leading
one-body terms is understood at least semi-quantitatively, as discussed below.
Two- and three-body contributions to the moments can be generated explicitly in
HBChPT by meson loop corrections to the leading one-body contributions to the
moments.  The loop corrections have been considered by many authors
\cite{Caldi,GSS,K,J,L,D,Mei,DH-ChPTmoments,DH-loop-moments}. Two- and three-body
contributions also appear through the nonzero initial values of the coefficients
$\mu_c$-$\mu_g$ which arise in dynamical models from Thomas-type terms associated
with short-distance QCD interactions and the long-range confining interaction
\cite{Durand}. A model which combines these calculations \cite{DH-loop-moments}
gives an excellent fit to the data, with a mean deviation of the calculated
moments from experiment of only 0.05 $\mu_N$. In the following sections, we discuss
the origin of the dynamical corrections to the QM moments, show that possible
orbital contributions to the moments are negligible, analyze how the loop and
dynamical corrections  combine to produce the good fit found in
\cite{DH-loop-moments}, and point out where problems remain.


\subsection{\label{subsec:background}Baryon moments in a QCD-based quark model}

\begin{figure}
\centerline{\hspace*{-2in}
\includegraphics{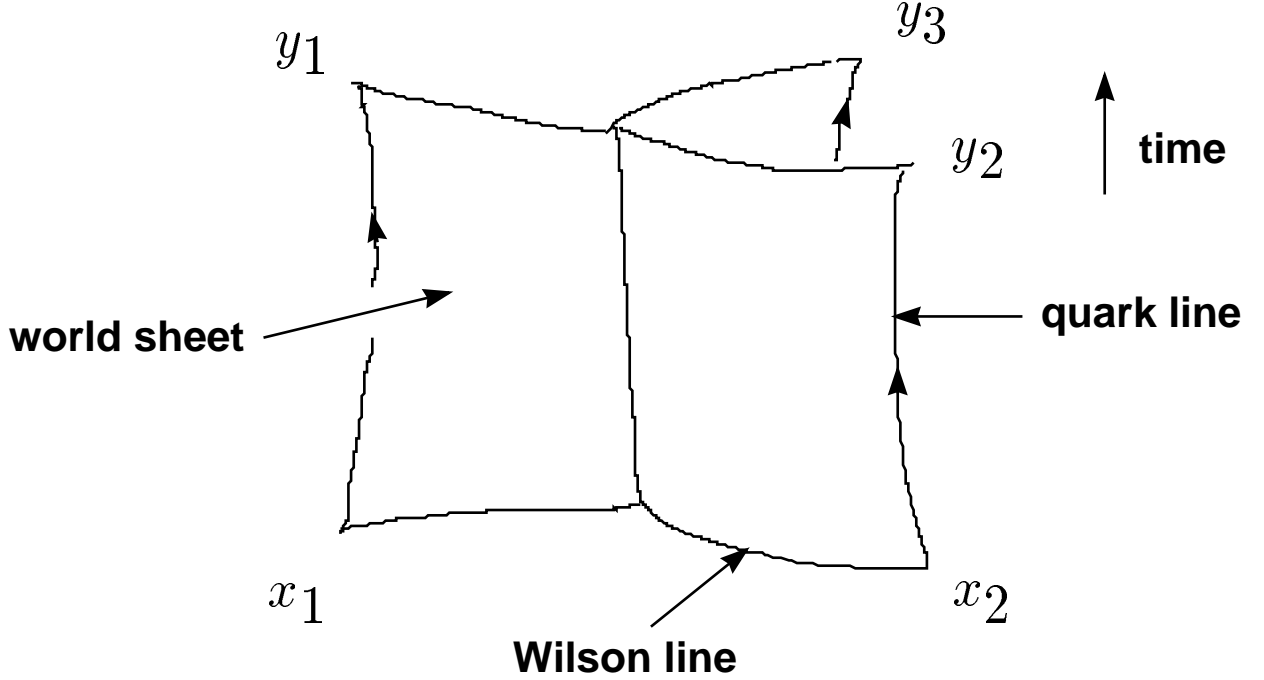}
}
\caption{\label{fig:worldsheet} World sheet picture for the structure of a baryon in
the Wilson-loop approach.}
\end{figure}

In previous work \cite{Durand,DH-loop-moments}, we derived the QM for the baryon
moments, including  dynamical corrections,
in the context of QCD. Our approach was based on the
work of Brambilla \textit{et al.} \cite{brambilla}, who derived the interaction
potential and wave equation for the valence quarks in a baryon
using a Wilson-line construction in quenched QCD.  Their basic idea was to construct a
Green's function for the propagation  of a gauge-invariant combination of
quarks joined by path ordered Wilson-line factors
\begin{equation}
U=P\exp\left( ig\!\int\! A_g\cdot dx\right), \label{wilsonline}
\end{equation}
where $A_g$ is the color gauge field. With internal quark loops omitted (the
quenched approximation to QCD), the Wilson lines sweep out a
three-sheeted world sheet of the form shown in Fig. \ref{fig:worldsheet}
as the quarks move from their initial to their final configurations. The
approximation ignores the effect of meson loops.


By making an expansion in powers of $1/m_j$  using the Foldy-Wouthuysen
approximation \cite{foldy}, and considering only forward
propagation of the quarks in time,
Brambilla \textit{et al.} were able to derive a Hamiltonian and Schr\"{o}dinger
equation for the quarks, with an interaction which involves an average over the
gauge field. That average was performed using the minimal surface approximation
in which fluctuations in the world sheet are ignored,
and the geometry is chosen to minimize the total area of the world sheet
subject to the motion of the quarks. The short-distance QCD
interactions were taken into account explicitly. Finally, the kinetic terms
could be resummed. The result of this construction was an effective Hamiltonian
\cite{brambilla} to be used in a semirelativistic Schr\"{o}dinger
equation $H\Psi=E\Psi$,
\begin{equation}
\label{hamiltonian}
H = \sum_i\sqrt{p_i^2+m_i^2}+\sigma(r_1+r_2+r_3)-\sum_{i<j}\frac{2}{3}
\frac{\alpha_{\textrm s}}{r_{ij}} + V_{SD},
\end{equation}
where $V_{SD}$ is the spin-dependent part of the potential.
Here ${\mathbf r}_{ij}=\bx_i-\bx_j$ is the
separation of quarks $i$ and $j$, $r_i$ is the distance of quark $i$ from
point at which the sum $r_1+r_2+r_3$ is minimized, the parameter $\sigma$
is a ``string tension'' which specifies the strength of the long range
confining interaction, and $\alpha_s$ is the strong coupling.

We will make one further approximation, known to be good \cite{kogut}, and will
replace $(r_1+r_2+r_3)$ in the confining potential by
$\frac{1}{2}(r_{12}+r_{23}+r_{31})$ and make corresponding changes in the
associated spin-dependent terms. This approximation allows simple analytic
calculation of a number of matrix elements we will need. With this change, the
Hamiltonian becomes
\begin{eqnarray}
H &=& H_0 + V_{SD},
\\
\label{H0}
H_0 &=& \sum_i\sqrt{{\pb}_i^2+m_i^2}
     +\frac{\sigma}{2}(r_{12}+r_{23}+r_{31})-\sum_{i<j}\frac{2}{3}
\frac{\alpha_{\rm s}}{r_{ij}},
\\
V_{SD} &=& -\frac{1}{4m_1^2}\frac{\sigma}{r_{12}}\bS_1\cdot({\rb}_{12} \times
{\pb}_1)-\frac{1}{4m_1^2}\frac{\sigma}{r_{13}}\bS_1\cdot({\rb}_{13} \times
{\pb}_1)  \nonumber \\
&& +\frac{1}{3m_1^2}\bS_1\cdot\left[({\rb}_{12}\times{\pb}_1)
\frac{\alpha_{\rm s}}{r_{12}^3} + ({\rb}_{13}\times{\pb}_1)
\frac{\alpha_{\rm s}}{r_{13}^3}\right] \nonumber\\
\label{V_SD}
&&-\frac{2}{3}\frac{1}{m_1m_2}\frac{\alpha_{\rm s}}{r_{12}^3}
\bS_1\cdot{\rb_{12}}\times{\pb}_2
-\frac{2}{3}\frac{1}{m_1m_3}\frac{\alpha_{\rm s}}{r_{13}^3}
\bS_1\cdot{\rb_{13}}\times{\pb}_3 \nonumber \\
&& +\frac{1}{m_1m_2}\frac{2}{3}\frac{\alpha_s}{r_{12}^3}\left[
\frac{3}{r_{12}^2}(\bS_1\cdot{\rb}_{12})(\bS_2\cdot{\rb}_{12})
- \bS_1\cdot\bS_2\right]  \\
&&+\frac{1}{m_1m_2}\frac{16\pi}{9}\alpha_s\delta^3({\rb}_{12})
\bS_1\cdot\bS_2+ \textrm{permutations}+\cdots, \nonumber
\end{eqnarray}
with $\bS_i=\bsig_i/2$  the spin operator for quark $i$.
The masses which appear in the spin-dependent terms are to be interpreted as effective
masses, with $1/m_i \simeq 1/E_i$, and are not necessarily equal to the masses with the
same labels that appear in $E_i=\sqrt{p_i^2+m_i^2}$. The  delta function spin-spin
interactions will not play a role in the following, and will be dropped. The terms
hidden in the ellipsis are momentum-dependent
interactions of ${\rm O}({\pb}^2/m^2)$ relative to the terms given explicitly. These
lead only to small corrections to the already-negligible orbital contributions to
the moments, and will be ignored. This model has been used in detailed calculations
in \cite{kogut} and \cite{isgur} to obtain good fits to the baryon spectrum up to
$\sim ~3$ GeV.

To derive a expression for the baryon moments in the same approximation to quenched
QCD, we redo the calculation of Brambilla \textit{et al.} with the gauge interaction
extended to include the electromagnetic vector potential
${\bA}_{\textrm em}(x_i)={\bf  B}\times{\bx}_q/2$ associated with a constant
external magnetic field $\textbf B$. We can then pick out the modified magnetic moment
operator through the relation
\begin{equation}
\Delta H=-\bmu\cdot{\textbf B} \label{deltaH}.
\end{equation}
The result is
\begin{equation}
\bmu=\sum_j \left(\bmu_j^{\rm (QM)} + \Delta \bmu_j^{QM}\right) + \bmu_L .
\end{equation}
Here $\bmu_j^{\rm (QM)}$ is the QM moment operator $\mu_j\bsig_j$, while
$\Delta \bmu_j^{QM}$ and $\bmu_L$ are the leading corrections to
the baryon moments associated with the binding interactions in $V_SD$ and
the nonzero orbital angular momentum in the baryon, respectively. The latter
two can be obtained directly by making the minimal substitution
${\pb}_i\longrightarrow {\pb}_i-e_i{\bA}({\bx}_i)$ in
Eq.\ (\ref{hamiltonian}) and isolating the $\textbf B$-dependent terms. This gives
\begin{eqnarray}
\Delta \bmu_1^{QM} &=& \frac{\mu_1}{6m_1}\frac{\alpha_s}{r_{12}^3}\left[
({\bx}_1\cdot{\rb}_{12})\bsig_1 -({\bx}_1\cdot\bsig_1){\rb}_{12}\right]
+(2\leftrightarrow 3) \nonumber
\\
\label{delta_mu_QM}
&+& \frac{\mu_2}{2m_1}\frac{\alpha_s}{r_{12}^3}\left[
({\bx}_2\cdot{\rb}_{21})\bsig_1 - ({\bx}_2\cdot\bsig_1) {\rb}_{21}\right]
+ (2\leftrightarrow 3) \nonumber
\\
&-& \frac{\mu_1}{4m_1}\frac{\sigma}{r_{12}}\left[
({\bx}_1\cdot{\rb}_{12})\bsig_1 - ({\bx}_1\cdot\bsig_1){\rb}_{12}\right]
+ (2\leftrightarrow 3),
\end{eqnarray}
where $\mu_i=e_i/2m_i$ with $m_i$ an effective mass. The results for
$\Delta \bmu_2^{QM}$ and $\Delta \bmu_3^{QM}$ which can be obtained by cyclic
permutation of the indices. Finally,
\begin{equation}
\bmu_L= \sum_j \mu_j{\bx}_j\times{\pb}_j = \bL_j
\end{equation}
where $\bL_j$ is the orbital angular
momentum of the $j$-th quark.

As discussed in \cite{DHJ2}, $\Delta \bmu^{QM}$ arises from
a set of Thomas precession terms. The diagonal terms in the first and third
lines of Eq.~(\ref{delta_mu_QM}) are
associated with the spin-same-orbit interaction in Eq.\ (\ref{V_SD}), and give
a largely multiplicative correction to $\mu_i$. The two-body spin-other-orbit
terms introduce new, nonadditive structure, and are important in improving the
simple quark-model fits to the moments \cite{DH-loop-moments}. These terms are
proportional to the short-distance spin-dependent part of the potential of
order $\alpha_s$ divided by $E_iE_j$, so are expected to be small.

To obtain an approximate operator form for the QM moment $\bmu_j$ to use in
later calculations, we act on the electromagnetic interaction term
$-e\balpha_i\cdot {\bA({\bx}_i)}$ with
the Foldy-Wouthuysen transformation which reduces the operator Dirac Hamiltonian
$(\balpha\cdot{\pb}+\beta m)$ for the free motion of a quark to  the
two-component form $H=\beta E$ \cite{foldy}, with $E =\sqrt{p^2+m^2}$  the
kinetic energy operator in Eq.~(\ref{H0}). This leads to O($e$) to a two-component
spin-dependent interaction
\begin{equation}
\label{reducedH_em}
-\frac{e_j}{2\sqrt{p_j^2+m_j^2}}\bsig\cdot{\textbf B} -
\frac{e_j}{4 \sqrt{p_j^2+m_j^2}(\sqrt{p_j^2+m_j^2}+m_j)}
\bsig\cdot({\textbf B}\times {\pb}) \times {\pb}.
\end{equation}
The first term can be identified with the QM moment interaction
$-\mu_j\bsig\cdot{\textbf B}$ and gives
\begin{equation}
\label{mu_0}
\mu_j=\left<e_j/(2E_j)\right>
\end{equation}
when averaged over the momentum distribution in a baryon. We will adopt this
form in later calculations to obtain reasonable estimates of the effect of the
different binding of the quarks in different baryons on the $\mu_j$, but will
otherwise treat the $\mu_j$ as parameters. In this treatment, the second
term in Eq.~(\ref{reducedH_em}), which behaves similarly as far as the averages
 are concerned, gives a single-particle contribution to the baryon moments which
 is absorbed in the adjustment of the parameters $\mu_j\simeq e_j/2m_j$ in terms
 of the effective masses.


\section{\label{sec:wavefuncs}BARYON WAVE FUNCTIONS}

\subsection{\label{subsec:wavefuncs}Spin structure of the baryon wave functions}

To estimate the dynamical corrections to the baryon moments, we need approximate
wave functions for the Hamiltonian in Eq.\ (\ref{hamiltonian}).
We have found it very useful to construct the wave functions for
the $j=\frac{1}{2}^+$ baryons in a form that allows us to use
trace methods to calculate the many spin matrix elements which
appear in the expressions for the magnetic moments. This
construction also has the advantage of displaying clearly the
symmetry properties and the internal angular momentum content of
the wave functions. The basic idea is simple. We start with the
$j=\frac{1}{2}$ spin wave functions for three quarks in the
quark-model ground state, and construct all further wave functions
by the application of even-parity scalar operators with the
correct symmetry to give allowed states. Because the total angular
momentum operator $\bf J$ commutes with scalar operators, the new
wave functions will retain the original eigenvalues $j(j+1)$,
$j_3$ of ${\bf J}^2$ and $J_3$. Our starting spin wave
functions are of the familiar form
\begin{eqnarray}
\chi_{\raisebox{-.5ex}{$\scriptstyle\frac{1}{2},\frac{1}{2}$}}^{S_{12}=1}
&=& \frac{1}{\sqrt{6}}[\,(\uparrow\downarrow+
\downarrow\uparrow)\uparrow-2(\uparrow\uparrow)\downarrow\,]\,,\nonumber\\
\chi_{\raisebox{-.5ex}{$\scriptstyle\frac{1}{2},-\frac{1}{2}$}}^{S_{12}=1}
&=& -\frac{1}{\sqrt{6}}[\,(\uparrow\downarrow+
\downarrow\uparrow)\downarrow-2(\downarrow\downarrow)\uparrow\,]\,,\nonumber\\
\chi_{\raisebox{-.5ex}{$\scriptstyle\frac{1}{2},\frac{1}{2}$}}^{S_{12}=0}
&=& \frac{1}{\sqrt{2}}(\uparrow\downarrow-
\downarrow\uparrow)\uparrow\,,\nonumber\\
\chi_{\raisebox{-.5ex}{$\scriptstyle\frac{1}{2},-\frac{1}{2}$}}^{S_{12}=0}
&=& \frac{1}{\sqrt{2}}(\uparrow\downarrow-
\downarrow\uparrow)\downarrow\,, \label{spinfuncs}
\end{eqnarray}
where we will label the quarks $1,\,2,\,3$ in order. The spin
functions for $S_{12}=0,\,1$ are connected by the operator
$(\bsig_1-\bsig_2)\cdot
\bsig_3$ which changes the symmetry in spins 1
and 2,
\begin{eqnarray}
\chi_{\raisebox{-.5ex}{$\scriptstyle\frac{1}{2},j_3$}}^{S_{12}=0}
&=&\frac{1}{2\sqrt{3}}(\bsig_1-\bsig_2)\cdot \bsig_3\,
\chi_{\raisebox{-.5ex}{$\scriptstyle\frac{1}{2},j_3$}}^{S_{12}=1},
\nonumber\\
\chi_{\raisebox{-.5ex}{$\scriptstyle\frac{1}{2},j_3$}}^{S_{12}=1}
&=&\frac{1}{\sqrt{2}}(\bsig_1-\bsig_2)\cdot \bsig_3\,
\chi_{\raisebox{-.5ex}{$\scriptstyle\frac{1}{2},j_3$}}^{S_{12}=0},
\label{chi1tochi0}
\end{eqnarray}
as is easily checked by direct calculation.

The functions $\chi_{\frac{1}{2},j_3}^{S_{12}=0,1}$ give a
complete basis for the $j=\frac{1}{2}$ spin states for three
quarks. A general $j=\frac{1}{2}$ wave function can therefore be
written in the form
\begin{eqnarray}
\psi_{\frac{1}{2},j_3} &=& |\mbox{$\frac{1}{2}$},j_3 \rangle =
\phi_1\,\chi_{\raisebox{-.5ex}
{$\scriptstyle\frac{1}{2},j_3$}}^{S_{12}=1}+\phi_0\,
\chi_{\raisebox{-.5ex}{$\scriptstyle\frac{1}{2},j_3$}}^{S_{12}=0}
\nonumber\\
&=&
\left[\,\phi_1+\phi_0\frac{1}{2\sqrt{3}}(\bsig_1
-\bsig_2)\cdot\bsig_3\,\right]\,
\chi_{\raisebox{-.5ex}{$\scriptstyle\frac{1}{2},j_3$}}^{S_{12}=1}
\equiv \widetilde{\phi}\,\chi_{\raisebox{-.5ex}
{$\scriptstyle\frac{1}{2},j_3$}}^{S_{12}=1}\, \label{psi}
\end{eqnarray}
where $\phi_1$ and $\phi_2$ are scalar operators symmetric in
$\bsig_1$ and $\bsig_2$. We
will construct the full operator $\widetilde{\phi}$ directly; the
part antisymmetric in $\bsig_1$ and
$\bsig_2$ will generate the $S_{12}=0$
component of $\psi$ when acting on the symmetrical $S_{12}=1$ spin
function.

We will concentrate initially on baryons containing two like
quarks, taken as 1 and 2 in our labeling. The singlet color wave
function of the baryon is antisymmetric in the interchange of any
two labels, so the space-spin wave function and, therefore, the
operator $\widetilde{\phi}$ must be symmetric under the
interchange of 1 and 2. No symmetry requirement exists with
respect to the third quark. We will work in the center-of-mass
system of the three quarks, and use the Jacobi coordinates
$\brho$ and $\blam$, defined
respectively as the separation of quarks 1 and 2, and the
separation of quark 3 from the center of mass of 1 and 2, to
describe the internal configuration of the baryon. It is also
useful to introduce Jacobi coordinates in which the pairs 2,3 and
3,1 are singled out, with
\begin{equation}
\begin{array}{cc}
\displaystyle\brho={\rb}_{12}, &
\displaystyle\blam={\rb}_{12,3},\\
\displaystyle\brho'={\rb}_{23}, &
\displaystyle\blam'={\rb}_{23,1},\\
\displaystyle\brho''={\rb}_{31}, &
\displaystyle\blam''={\rb}_{31,2},
\end{array} \label{jacobi_coord}
\end{equation}
where ${\rb}_{ij}={\bz}_i-{\bz}_j$, ${\rb}_{ij,k}={\bf
R}_{ij}- {\bz}_k$, and ${\bf R}_{ij}=(m_i{\bz}_i+m_j{\bz}_j)/m_{ij}$. The
relations among these coordinates and among the
corresponding momenta are given in Appendix \ref{app:jacobi}. We
note here that $\brho$ and $\blam$ are, respectively, odd and even under
the the interchange of 1 and 2.

It will be useful before determining $\widetilde{\phi}$ to
determine the possible orbital angular momentum configurations for
a spin $j=\frac{1}{2}^+$ baryon. We introduce two angular momenta,
$L_\rho$ between quarks 1 and 2, and $L_\lambda$ between quark 3
and the center of mass of 1 and 2, and combine these to produce
the total orbital angular momentum $L$. This must combine in turn
with the total quark spin $\bS=\frac{1}{2},\, \frac{3}{2}$ to
give $j=\frac{1}{2}^+$. Clearly, $L\leq 2$, while $L_\rho$ and
$L_\lambda$ must both be even or both odd to give a positive
parity state. If we restrict our attention to states with
$L_\rho,\,L_\lambda, L_\rho+L_\lambda\leq 2$ as well, the complete
wave function can be constructed using the configurations
\begin{equation}
(L_\rho,L_\lambda,L)=\left\{
\begin{array}{lll}
(0,0,0) & (1,1,0) & (1,1,1) \\
(1,1,2) & (2,0,2) & (0,2,2)
\end{array}
\right. \label{states}
\end{equation}
Each unit of $L_\rho$ ($L_\lambda$) in the wave function
corresponds to one factor of $\brho$
($\blam$) in the corresponding angular momentum
tensor, and a symmetry factor -1 (+1) for an interchange of quarks
1 and 2. The required symmetry of the wave function under quark
exchange then limits $S_{12}$ to the values $S_{12}=1$
($S_{12}=0$) for $L_\rho$ even (odd) for the baryons with quarks 1
and 2 identical. This eliminates the configuration (1,1,2),
$S_{12}=0$ since this combination of angular momenta gives
$j=3/2,\,5/2$ only.

It is now straightforward to construct the operator
$\widetilde{\phi}$ in Eq.\ (\ref{psi}) using the Pauli spin
matrices $\bsig_i$, the coordinate vectors
$\brho$, $\blam$, and the tensor
$t_{ij}$,
\begin{equation}
\label{tij} t_{i,j}(\bx)=3\bsig_i\cdot\!
\bx\,\, \bsig_j\!\cdot\!\bx-\bsig_i\!\cdot\!\bsig_j\,\bx^2.
\end{equation}

The possible components are given in the labelling
$(L_\rho,L_\lambda,L)S_{12}$ by
\begin{eqnarray} \label{allstates}
(0,0,0)1: && 1 \\
(1,1,0)0: &&
(\bsig_1-\bsig_2)\cdot
\bsig_3\,\brho\cdot
\blam \\
\label{row3}
(1,1,1)0: &&
(\bsig_1-\bsig_2)\cdot
\brho\times\blam,\quad
(\bsig_1-\bsig_2)\times
\bsig_3\cdot(\brho\times
\blam)\\
\label{row4}
(2,0,2)1: && t_{12}(\brho),\quad
t_{23}(\brho)
+t_{31}(\brho) \\
\label{row5}
(0,2,2)1: && t_{12}(\blam),\quad t_{23}
(\blam)+t_{31}(\blam)
\end{eqnarray}
Two operators are shown in the rows for $L=1,\, 2$. However, it is
clear that the two cannot be independent for the $j=\frac{1}{2}^+$
baryons because there is only one way to reach $j=\frac{1}{2}$ by
combining the indicated values of $L$, $S_{12}$, and
$s_3=\frac{1}{2}$. We show directly in Appendix \ref{app:traces}
that the second operators are  not independent of the first for
$j=\frac{1}{2}$  (this is not the case for $L=2,\,j=\frac{3}
{2}$). We will therefore write $\widetilde{\phi}$ in terms of the
first operator in each row, with
\begin{equation}
\widetilde{\phi}=a+ib\,(\bsig_1-\bsig_2)
\cdot\brho\times\blam
+c\,t_{12}(\brho)+d\,t_{12}(\blam)
+e\,(\bsig_1-\bsig_2)\cdot
\bsig_3\,\brho\cdot
\blam,
\label{wavefunc}
\end{equation}
where $a,\dots,e$ are functions of $\rho^2$ and $\lambda^2$
only \footnote{In the general case with one or both of the internal
angular momenta $L_\rho,\,L_\lambda$ greater than two, there are
additional tensors with the opposite 1,2 symmetry which appear
multiplied by a factor $(\brho\cdot
\blam)$ to restore the overall symmetry of
$\widetilde{\phi}$, and the coefficient functions depend on
$(\brho\cdot \blam)^2$ as
well as $\rho^2$ and $\lambda^2$. See the following discussion of
the opposite symmetry tensors in connection with the $\Lambda$
hyperon. }. In the general case, the coefficient functions can
also depend on $(\brho\cdot\blam)^2$, but this requires the introduction
of values of $L_\rho=L_\lambda\geq 2$.

The analysis above holds for quarks 1 and 2 identical, that is, for the
$p$, $n$, $\Sigma^\pm$, $\Xi^0$, and $\Xi^-$ baryons. The $\Sigma^0$ and
$\Lambda$ are both $uds$ systems, and are distinguished by the interchange
symmetry of the $ud$ pair. This is even for the $\Sigma^0$, so the
$\Sigma^0$ wave function has the structure given in Eq.\ (\ref{wavefunc}).
The $\Lambda$ is different, with odd $ud$ interchange symmetry. An analysis
similar to that above gives the spin dependence of the $\Lambda$ wave function
as $\psi_{\frac{1}{2},j_3}^\Lambda = \widetilde{\phi}_\Lambda
\chi_{\frac{1}{2},j_3}^{S_{12}=1}$, with
\begin{eqnarray}
\widetilde{\phi}_\Lambda& &= a_\Lambda\bsig_3\cdot(\bsig_1-\bsig_2) +
b_\Lambda\brho\cdot\blam +  c_\Lambda i\bsig_3\cdot\brho\times\blam \nonumber \\
\label{lambda_wf}
&& + d_\Lambda \brho\times\blam\cdot\bsig_3\times(\bsig_1+\bsig_2) +
e_\Lambda t_{12}(\brho,\blam),
\end{eqnarray}
where
\begin{equation}
\label{t12(rho,lambda)}
t_{12}(\brho,\blam)=\frac{1}{2}[\,3\bsig_1\!\cdot\!\brho\,
\bsig_2\!\cdot\!\blam - \bsig_1\!\cdot\!\bsig_2 \,\brho\!\cdot\!\blam +
(1\leftrightarrow 2)].
\end{equation}
%


\subsection{\label{subsec:traces}Trace methods}

Since the total angular momentum operator $\bf J$ commutes with the Hamiltonian
$H$, the matrix elements of $H$ satisfy the relations

\begin{eqnarray}
\langle j',j_3' | H | j, j_3 \rangle &=&
\langle j',j_3' | H | j, j_3 \rangle \delta_{j'j}\delta_{j_3'j_3} ,
\\
\langle j,j_3' | H | j, j_3' \rangle &=&
\langle j,j_3 | H | j, j_3 \rangle ,
\end{eqnarray}
so, for $j=\frac{1}{2}$, $j_3=\pm \frac{1}{2}$
\begin{eqnarray}
\langle \mbox{$\frac{1}{2}$},j_3 | H | \mbox{$\frac{1}{2}$}, j_3 \rangle
& = & \frac{1}{2}[\langle \mbox{$\frac{1}{2}$}, \mbox{$\frac{1}{2}$} | H |
\mbox{$\frac{1}{2}$}, \mbox{$\frac{1}{2}$} \rangle
+ \langle \mbox{$\frac{1}{2}$}, \mbox{-$\frac{1}{2}$}
| H | \mbox{$\frac{1}{2}$}, \mbox{-$\frac{1}{2}$} \rangle ] \nonumber \\
 & = & \frac{1}{2} {\rm Tr}_{j=\mbox{$\frac{1}{2}$}} H \, .
\end{eqnarray}
We can use the projection operators to get rid of the restriction to $j=\frac{1}{2}$
in the trace.  Formally,
\begin{equation}
P_{S, S_{12}}= \sum_{j_3} | S_{12}; s, j_3 \rangle
               \langle S_{12}; s, j_3 | \, .
\end{equation}
Now $P_{\frac{1}{2},1}\chi_{j3}^{S_{12}=1}= \chi_{j_3}^{S_{12}=1}$, while
$P_{\frac{1}{2},1}$ annihilates $\psi_{\frac{3}{2},j_3}$, so by introducing the
projection operator
\begin{equation}
P_{\frac{1}{2},1}=\frac{1}{12}[ 3 +\bsig_1\cdot
\bsig_2- 2\bsig_1\cdot \bsig_3 - 2\bsig_2\cdot \bsig_3]
\end{equation}
in matrix elements, we can use a basis of independent states $\uparrow$,
$\downarrow$ for quarks 1,2, and 3 and sum over the spin indices without
restriction. This gives
\begin{eqnarray}
\langle \psi_{\frac{1}{2},j_3} | H | \phi_{\frac{1}{2},j_3} \rangle
&=&\frac{1}{2} \sum_{j_3=\pm \frac{1}{2}}
\chi_{\raisebox{-.5ex}
{$\scriptstyle\frac{1}{2},j_3$}}^{\dagger S_{12}=1}\widetilde{\psi}^{\dagger}
H \widetilde{\phi}\chi_{\raisebox{-.5ex}
{$\scriptstyle\frac{1}{2},j_3$}}^{S_{12}=1} \nonumber \\
&=&\frac{1}{2} \sum_{\lambda_1, \lambda_2, \lambda_3 =\pm \frac{1}{2}}
\widetilde{\psi}^{\dagger}
H \widetilde{\phi}\, P_{\frac{1}{2},1} \nonumber \\
&=&\frac{1}{2} \, {\rm Tr} \widetilde{\psi}^{\dagger}
H \widetilde{\phi}\, P_{\frac{1}{2},1}
\end{eqnarray}
where ${\rm Tr}={\rm Tr}_1 {\rm Tr}_2 {\rm Tr}_3$ and $\lambda_i$
is the spin projection of the $i$-th quark ($i=1,2,3$). Note that
\begin{eqnarray}
{\rm Tr} \, \openone &=& 8 \, , \nonumber \\
\frac{1}{2} {\rm Tr}\, P_{\frac{1}{2},1} &=& 1 \, , \\
{\rm Tr} \, \bsig_1 = {\rm Tr} \, \bsig_2
= {\rm Tr} \, \bsig_3 &=& 0 \, . \nonumber
\end{eqnarray}


\subsection{Approximate wavefunctions}

To get approximate wave functions to use in our calculations of the binding
corrections to the moments,
we have made variational calculations of the energies and approximate
wave functions of the ground state baryons and their first excited states
using the Hamiltonian in Eqs.~(\ref{hamiltonian}) and (\ref{V_SD}). Our calculation
and results are summarized below.
Some details are given in Appendix A.

We emphasize that we are not trying to make detailed, accurate calculations, but
rather to obtain crude wave functions which allow us to estimate the small binding
corrections. More accurate calculations would probably only be useful in conjunction
with further improvements in the underlying theory.

We rewrite the baryon wave function in Eq. (\ref{wavefunc}) as
\begin{equation}
\label{completewf}
\psi_{\frac{1}{2},m} = (\psi_0^a + b_0 {\hat\psi}_b + c_0 {\hat\psi}_c +
     d_0 {\hat\psi}_d + e_0 {\hat\psi}_e ) \chi_{\raisebox{-.5ex}
{$\scriptstyle\frac{1}{2},m$}}^{S_{12}=1} \, ,
\end{equation}
where $\psi_0^a$ is a normalized wave function of the ground state and
\begin{eqnarray}
\hat\psi_b &=& iN_b\, (\bsig_1-\bsig_2)
\cdot\brho\times\blam\,\psi_0^b
\nonumber\\
\hat\psi_c &=& N_c\, t_{12}(\brho)\,\psi_0^c
\\ \label{wavefunc_re}
\hat\psi_d &=& N_d\, t_{12}(\blam)\,\psi_0^d
\nonumber\\
\hat\psi_e &=& N_e\, (\bsig_1-\bsig_2)\cdot
\bsig_3\,\brho\cdot
\blam\,\psi_0^e \, ,
\nonumber
\end{eqnarray}
are the normalized wave functions of the excited states. Now if we choose
$\psi_0^a, \cdots, \psi_0^e$ as the Gaussians \footnote{We acually considered
somewhat more general functions with Gaussians multiplied by polynomials, but the
basic results changed rather little. The simple Gaussians are sufficient to
illustrate the main features of the corrections.},
\begin{equation}
\label{gausswf}
\psi_0^A = \left( \frac {\beta_\rho^A \beta_\lambda^A }{ \pi} \right)^{3/2}
 \exp{ [ -({\beta_\rho^A}^2 \rho^2 + {\beta_\lambda^A}^2 \lambda^2)]} \, ,
\ \ A = a, \cdots, e \, ,
\end{equation}
then the normalization coeffcients $N_b \,, \cdots, N_e$ are
\begin{equation}
N_b = \frac {\beta_\rho^b \beta_\lambda^b}{\sqrt{2}} \, , \ \
N_c = \frac {{\beta_\rho^c}^2}{\sqrt{30}} \, , \ \
N_d = \frac {{\beta_\lambda^d}^2}{\sqrt{30}} \, , \ \
N_e = \frac {\beta_\rho^e \beta_\lambda^e}{ 3} \, .
\end{equation}

The coefficients $b_0 \,, \cdots, e_0$ can be evaluated using perturbation
theory in which the excited states are considered as the perturbation states.
In such a way, we have
\begin{equation}
A_0 = \frac{ \langle {\hat\psi}_A | V_{SD} | \psi_0^a \rangle}{E_0 - E_A }
 \, , \ \ A = b, \cdots, e \, ,
\end{equation}
where $V_{SD}$ is the spin-dependent potential in
Eq. (\ref{V_SD}) with the spin-spin and higher order terms omitted, and
\begin{equation}
E_A =  \langle {\hat\psi}_A | H_0 | {\hat\psi}_A \rangle
 \, , \ \ E_0 =  \langle \psi_0^a | H_0 | \psi_0^a\rangle  .
\end{equation}
$H_0$ is given in Eq.~(\ref{H0}).

To evaluate the coefficients $b_0 \,, \cdots, e_0$ numerically, we first minimize
the energies $E_A$ and $E_0$ by varying the parameters
$\beta_\rho$, $\beta_\lambda$, and thus obtain the best variational wave
functions $\psi_0^a$, ${\hat\psi}_b$, ..., ${\hat\psi}_e$ of Gaussian form.
The calculation uses
$\alpha_s=0.39$, $\sigma=0.18 \, {\rm GeV}^2$, $m_u=m_d=0.200$ GeV,
and $m_s=0.500$ GeV. The resulting values of $\beta_\rho$ and $\beta_\lambda$ are
given in Table\ \ref{table:beta_parameters}. These values differ for different
baryons, a point which will be important later. The values of excitation energies
$E_A-E_0$ obtained this way and shown  in Table \ref{table:energy_diff} are
generally consistent with those found in the more elaborate calculations
\cite{isgur} \footnote{ Note that here we have not considered the first excited
state of $L=0$.}.

These results together with those from our
calculation of the matrix elements $\langle {\hat\psi}_A | V_{SD} | \psi_0^a \rangle $
between the ground state and lowest excited states $b \,, \cdots, e$
allow us to estimate the coefficients $b_0 \,, \cdots, e_0$.  We  list the
values found in Table \ref{table:wfcoeff}.


%
\begin{table}[htbp]
\centering
\caption{ The energy differences $ \Delta E_A$ in MeV between the baryon ground states
$\psi_0$ and the excited states $\hat{\psi}_A$, $A=b,\ldots,\,e$ in
Eq.\ (\protect\ref{wavefunc_re}).}
\label{table:energy_diff}
\medskip
\begin{ruledtabular}
\begin{tabular}{crrrr}
     Baryon & $ \Delta E_b$ & $ \Delta E_c$ & $ \Delta E_d$ & $ \Delta E_e$  \\
    \hline
    $N$      & 870 & 821 & 820 & 789 \\
    $\Sigma$ & 841 & 837 & 750 & 755 \\
    $\Xi$    & 810 & 714 & 806 & 738 \\
    $\Omega$ & 779 & 732 & 730 & 701 \\
\end{tabular}
\end{ruledtabular}
\end{table}
\begin{table}
\caption{Values of the parameters $\beta_\rho$ and $\beta_\lambda$ in the Gaussian
approximation to the ground state wave functions of the octet baryons,
Eq.\ (\protect\ref{gausswf}), calculated for $m_u=m_d=0.200$ GeV,
$m_s=0.500$ GeV, $\alpha_s=0.39$, and $\sigma=0.18$ GeV$^2$.}
\label{table:beta_parameters}
\begin{ruledtabular}
\begin{tabular}{ccccc}
Baryon & $N$ & $\Sigma$ & $\Xi$ & $\Omega$ \\
\hline
$\beta_\rho$ & 0.340 & 0.347 & 0.387 & 0.394 \\
$\beta_\lambda$ & 0.393 & 0.424 & 0.420 & 0.455
\end{tabular}
\end{ruledtabular}
\end{table}
\begin{table}[htbp]
\centering
\caption{ The coefficients $b_0,\ldots, e_0$ are evaluated using
perturbation theory for
$\alpha_s=0.39$, $\sigma=0.18 {\rm GeV}^2$, $m_u=m_d=0.343$ GeV,
and $m_s=0.539$ GeV. }
\label{table:wfcoeff}
\begin{ruledtabular}
\begin{tabular}{crrrr}
     Baryon & $b_0$ & $c_0$ & $d_0$ & $e_0$  \\
    \hline
    $N$      & 0 & -0.013 & 0.013 & -0.021 \\
    $\Sigma$ & 0.012 & -0.013 & 0.011 & -0.016 \\
    $\Xi$    & -0.006 & -0.009 & 0.011 & -0.020 \\
    $\Omega$ & 0 & -0.009 & 0.010 & -0.016 \\
\end{tabular}
\end{ruledtabular}
\end{table}
%


While the Hamiltonian does mix orbitally excited states into the
$L_\rho=L_\lambda=L=0$ quark-model ground state,
the coefficients are very small, ranging from essentially zero to about
0.02 depending on the baryon. Because these coefficients only
appear quadratically in the baryon moments, the orbital contributions to the
moments will be completely negligible. This result is consistent with the
Lichtenberg's finding for the nonrelativistic QM \cite{Lichtenberg}.


\section{\label{sec:calculation}ANALYSIS OF THE BARYON MOMENTS}

\subsection{\label{subsec:leading}The leading approximation to the moments}

The leading approximation to the QM moments in Eq.~(\ref{mu_0}) gives
$\mu_j=\langle e_j/E_j\rangle=\langle e_j/(2\sqrt{p^2+m_j^2}) \rangle$.
The matrix elements $\langle 1/E_j \rangle$, customarily written in terms
of an effective mass as $e_j/m_j$, will actually differ for a given quark
depending on the baryon in which it is confined. To estimate this effect,
we have calculated the relevant matrix elements using the Gaussian variational
wave functions and the methods sketched in Appendix \ref{app:matrix_elements}.
The calculation was done using $m_u=m_d=0.200$ GeV and $m_s=0.500$ GeV in $H_0$,
values which give both a good ground-state baryon spectrum and reasonable values
for the moments. We then estimate the corrections to $\mu_u$ by taking the ratios
of the calculated matrix elements in the $\Sigma^+$, $\Lambda$, and $\Xi^0$ to
the matrix element in the proton, and multiplying by the actual value
$\mu_u=2\mu_a/3$ determined from the data. The correction to $\mu_s$ in the $\Xi$
states is determined similarly using the $\Sigma$ as the reference state. The
corrections are small, with, for example, $\mu_u$ decreasing by 0.046 $\mu_N$ in
the $\Sigma$ and 0.088 $\mu_N$ in the $\Xi$ relative to its value in $N$.

The overall corrections to the parameters $\mu_a,\ldots,\mu_{MM}$ are given in
Table~\ref{table:mu_parameters}.  The largest correction is to $\mu_d$. This
arises, as suggested by Fig.~\ref{fig1}\,(d) from the change in the moment of
one quark, relative to the values for the $m_i$ given above, which arises from
a change in the wave function when a second quark is strange.

The nonzero value of $\mu_{MM}$ found in this calculation corresponds to the
introduction of the new structure
\begin{equation}
\label{mB_MM}
{\mb}_{B,MM} = \sum_{i\not=j\not=k}\left[Q_iM_jM_k-\openone_i(QM)_jM_k\right] \bsig
\end{equation}
at the baryon level, where $\bsig$ is twice the baryon spin operator. This is a
mixed two- and three-body operator. The operator coefficient of $\bsig$ vanishes
for all states which do not contain two strange quarks, and has the value 1 for
the $\Xi^0$ and 0 for the $\Xi^-$. There are no further independent structures
for the octet baryons \footnote{As was noted in \cite{DH-loop-moments}, any
correction which affects only the $\Xi^-$ can be absorbed through an adjustment
of the other parameters.}.


\subsection{\label{subsec:binding}Binding corrections}

Without the corrections from the internal orbital angular moments,
the moment of a baryon $B$ is given by
\begin{equation}
\mu_B=\sum_{j}(\mu_j+\Delta\mu_j^B)\langle\sigma_{j,z}\rangle_B
=\mu_B^{QM}+\sum_j\Delta\mu_j^B\langle\sigma_{j,z}\rangle_B,
\label{QMmufinal}
\end{equation}
where the sum is over the quarks in the baryon and
we have quantized along $\textbf B$, taken along the $z$ axis. The spin
expectation values are to be calculated in the baryon ground state.
The correction $\Delta\mu_j^B$ to the  moment of quark $j$
depends on the baryon $B$ in which it appears. The final baryon moments depart
from the quark model pattern only when
the ratios $\Delta\mu_j^B/\mu_j$ differ in different baryons.

The general result for the moment operator given above
can be simplified considerably for the $L=0$
ground state baryons. The absence of any Pauli matrices in the wave functions
allows us to reduce the operators in $\Delta\mu_i^{QM}$ in Eq.\ (\ref{delta_mu_QM})
to the components proportional to $\bsig_i$ in calculating the ground-state matrix
elements,
\begin{equation}
\label{mu_projection}
\langle \Delta\bmu_i^{QM}\rangle_B = \langle \Delta\mu_i^{QM}\bsig_i
\rangle_B \, , \ \  \Delta\mu_i^{QM}=\frac{1}{6}{\rm Tr}\,\bsig_i\cdot\Delta\bmu_i^{QM},
\end{equation}
and the contribution to the quark moment $\mu_i$ in baryon $B$ is just
$\Delta\mu_i^B=\langle\Delta\mu_i^{QM}\rangle_B$. We will write $\Delta\mu_i^B$
as the sum of terms arising from diagonal and Thomas-type corrections,
\begin{equation}
\label{delta,lambda}
\Delta\mu_i^B = \Delta\mu_i^D + \Delta\mu_i^T.
\end{equation}
These are given by
\begin{equation}
\label{delta_mu_i}
\Delta\mu_i^D = \frac{\mu_i}{2m_i}\sum_{j\not=i}\left(\epsilon_{ij}
-\Sigma_{ij}\right),\ \  \Delta\mu_i^T =
\frac{\mu_i}{e_i}\sum_{j\not=i}\frac{e_j}{m_j}\epsilon_{ji},
\end{equation}
where the $\epsilon$'s and $\Sigma$'s are the ground state matrix elements

\begin{equation}
\label{epsilon_ij}
\epsilon_{ij} = \frac{2\alpha_s}{9}\Big\langle\frac{{\rb}_{ij}
\cdot{\bx}_j}{r_{ij}^3}\Big\rangle_B \, , \ \  \Sigma_{ij} =
\frac{\sigma}{6}\Big\langle\frac{{\rb}_{ij}\cdot\bx_i}{r_{ij}}
\Big\rangle_{_B}.
\end{equation}

Two quarks always have the same mass in each octet or decuplet baryon,
so appear symmetrically in the spatial part of a flavor-independent
wave function. We will label these quarks 1 and 2, let $m_1=m_2=m$,
and take 3 as the odd-mass quark if there is one.
With this labeling, the spatial matrix elements
can be reduced to the small set
\begin{eqnarray}
\label{epsilons}
\epsilon = \epsilon_{12} = \epsilon_{21}, \qquad
\epsilon' &=& \epsilon_{13} = \epsilon_{23}, \qquad
\tilde{\epsilon} = \epsilon_{31} = \epsilon_{32}, \\
\label{sigmas}
\Sigma = \Sigma_{12} = \Sigma_{21}, \qquad \Sigma' &=& \Sigma_{13} =
\Sigma_{23}, \qquad \widetilde{\Sigma} = \Sigma_{31} = \Sigma_{32},
\end{eqnarray}
 and the diagonal correction terms become
\begin{eqnarray}
\label{delta_mu^D}
\Delta\mu_1^D&=&\frac{\mu_1}{2m}\left(\epsilon+\epsilon'-\Sigma-
\Sigma' \right), \nonumber \\
\Delta\mu_2^D&=& \frac{\mu_2}{2m}\left(\epsilon+\epsilon'
-\Sigma-\Sigma' \right), \nonumber \\
\Delta\mu_3^D &=& \frac{\mu_3}{m_3} \left(\tilde{\epsilon}-\widetilde{ \Sigma}\right).
\end{eqnarray}
The Thomas-type corrections are somewhat more complicated,
\begin{eqnarray}
\label{delta_mu^T}
\Delta\mu_1^T &=& \frac{\mu_1}{e_1}\left(
\frac{e_2}{m}\epsilon+\frac{e_3}{m_3}\tilde{\epsilon}\right), \nonumber \\
\Delta\mu_2^T &=& \frac{\mu_2}{e_2}\left(
\frac{e_1}{m}\epsilon+\frac{e_3}{m_3}\tilde{\epsilon}\right), \nonumber \\
\Delta\mu_3^T &=&\frac{\mu_3}{e_3} \frac{e_1+e_2}{m}\epsilon'.
\end{eqnarray}
Note that, for all masses equal, $\epsilon=\epsilon'=\tilde{\epsilon}$,
and $\Sigma=\Sigma'=\widetilde{\Sigma}$. The matrix elements differ only
because of effects of the odd-quark mass on the wave functions.


%
\begin{table}
\caption{ The values in GeV of the matrix elements $\epsilon$ and $\Sigma$
defined in Eqs. (\protect\ref{epsilons}) and (\protect\ref{sigmas})
The matrix elements were evaluated for
$\alpha_s=0.39$, $\sigma=0.183\ {\rm GeV}^2$, $m_u=m_d=0.200$ GeV, and
$m_s=0.500$ GeV for the quark moments $\mu_u=1.943$, $\mu_s=-0.640$ obtained
from a fit to the data. }
\label{table:matrix_elements}

\begin{ruledtabular}
\begin{tabular}{crrrrrr}
     Baryon & $\epsilon$ & $\epsilon'$ & $\tilde{\epsilon}$
       & $\Sigma$ & $\Sigma'$ & $\widetilde{\Sigma}$  \\
    \hline
    $N$      & 0.017 & 0.017 & 0.017 & 0.050 & 0.050 &  0.050 \\
    $\Sigma$ & 0.017 & 0.024 & 0.011 & 0.049 & 0.064 &  0.031 \\
    $\Xi$    & 0.019 & 0.013 & 0.023 & 0.044 & 0.033 &  0.060 \\
    $\Omega$ & 0.019 & 0.019 & 0.019 & 0.044 & 0.044 &  0.044
\end{tabular}
\end{ruledtabular}
\end{table}
%

We have evaluated the radial matrix elements above using the Gaussian
wave functions obtained in our variational calculation of the ground
state energies for the Hamiltonian in Eq.\ (\ref{hamiltonian}) and the
methods sketched in Appendix \ref{app:matrix_elements}. The results
are given in Table \ref{table:matrix_elements} for
$\alpha_s=0.39$ and $\sigma=0.18\ {\rm GeV}^2$, values taken from
fits to the baryon spectrum \cite{kogut,isgur} using the same Hamiltonian.
As is clear from Table \ref{table:matrix_elements} the $\epsilon$'s
and the $\Sigma$'s are all similar in magnitude even for $m_1\not=m_3$, so
the main effect of the diagonal corrections $\Delta\mu_i^D$ in
Eq.~(\ref{delta_mu^D}) is a uniform shift in the input values of the
$\mu_i$'s. This is absorbed when $\mu_u=-2\mu_d$ and $\mu_s$ are used as
parameters in fitting the observed moments. The only nontrivial two- or
three-body effect of these terms arises from the differences between the
matrix elements in the different baryons. In contrast, the Thomas-type
corrections $\Delta\mu_i^T$ in Eq.\ (\ref{delta_mu^T}) are mainly two-body
contributions and remain nontrivial even when the $\epsilon$'s are similar
because of the different charges of the different quarks.

%
\begin{table*}
\caption{The parameters $\mu_a$ - $\mu_g$ and $\mu_{MM}$ for the estimated binding
corrections to the moments, the diagonal and  Thomas-type corrections defined in
Eqs.\ (\protect\ref{delta_mu^D}) and (\protect\ref{delta_mu^T}), and for the meson loop
corrections. The last lines give the total correction and the values of the parameters
obtained in a least squares fit to the measured moments, and the differences with the
experimental uncertainties.
}
\begin{ruledtabular}
\begin{tabular}{ccccccccc}
Type & $\mu_a$ & $\mu_b$ & $\mu_c$ & $\mu_d$ & $\mu_e$ & $\mu_f$ & $\mu_g$ & $\mu_{MM}$ \\
\hline
$\Delta\mu^0$ & 0 & 0.011 & 0 & -0.067 & 0.011 & 0.003 & 0.003 & -0.019  \\
$\Delta\mu^D$ & -0.306 & 0.231 & 0.000 & -0.023 &  0.001 & 0.000 & -0.008 & -0.012\\
$\Delta\mu^T$ & 0 &  0.000 & 0.151 & 0.000 & -0.006 & -0.082 & 0.004 & -0.013\\
meson loops & 0.123 & -0.208 & -0.339 & -0.022 & 0.143 & 0.471 & 0.185& 0.010 \\
\hline
$\Delta\mu_{\rm total}$ &  -0.183 & 0.034 & -0.188 & -0.112 & 0.149 & 0.392 & 0.184 & -0.034\\
Fit to data & 2.793 & -0.933 &-0.077 & -0.037 & 0.098 & 0.438 & 0.044 & 221 \\
& & $\pm 0.012$ & & $\pm 0.070$ & $\pm 0.010$ & $\pm 0.036$ & $\pm 0.070$ & $\pm 0.140$ \\
\hline
${\rm Fit}-\Delta\mu_{\rm total}$ & 2.976 & -0.899 & 0.111 &-0.075 & -0.051 & 0.046
& -0.100 & 0.255 \\
& & $\pm 0.012$ & & $\pm 0.070$ & $\pm 0.010$ & $\pm 0.036$ & $\pm 0.070$ & $\pm 0.140$
\end{tabular}
\label{table:mu_parameters}
\end{ruledtabular}
\end{table*}
%

In Table \ref{table:mu_parameters} we shown the values of the parameters
$\mu_a$ - $\mu_g,\,\mu_{MM}$ for the  diagonal and Thomas-type corrections
obtained using  the input moments $\mu_u=1.94$ $\mu_N$ and $\mu_s=-0.64$ $\mu_N$ found
in a fit to the data which includes these and the other corrections. The
effective masses used in Eqs.~(\ref{delta_mu^D}) and (\ref{delta_mu^T}) were
defined in terms of the input moments since the original coefficients in
Eq.~(\ref{delta_mu_QM}) are of the form $e_i/m_i^2$ or $e_i/m_im_j$.  The
smallness of the parameters $\mu_c$ - $\mu_g$ for the diagonal corrections
compared to the values of $\mu_a$ and $\mu_b$ shows that the $\Delta\mu_i^D$
are indeed mostly contributions from the one- body operators. In contrast,
the $\Delta\mu_i^T$ contribute mostly to the two-body coefficients
$\mu_c$ - $\mu_g$ with the largest contribution for the Thomas terms
$\mu_c$ and $\mu_f$, Fig.~\ref{fig1}. The value of $\mu_g$, which is associated
with the three- body operators, is very small is both cases. Finally, the
nonzero values of $\mu_{MM}$ obtained in both cases arise indirectly from the
dependence of the matrix elements in Eqs.~(\ref{epsilons}) and (\ref{sigmas})
on the strange-quark content of the state.


\subsection{\label{subsec:L>0}Contributions of internal orbital angular momenta}

As already remarked, the contributions of orbital angular momenta to the moments
are negligible for the ground-state octet baryons. We will illustrate this in the
special case in which  $m_1=m_2$ and $\mu_1=\mu_2=m$. $\bmu_L$ is given in this case by
\begin{equation}\label{mm_l}
\bmu_L=\mu_1 \bL_\rho + \frac{1}{M}\left(m_3 \mu_1 + 2m \mu_3 \right)
\bL_\lambda \ ,
\end{equation}
where $\bL_\rho=\brho \times {\pb}_\rho$,
$\bL_\lambda=\blam \times {\pb}_\lambda$, and
$M=2m_1+m_3$.

Using the wave function given in Eq.\ (\ref{completewf}) and choosing
the Gaussian spatial wave functions discussed above, we calculate
$\langle \bmu \rangle$ and obtain for the moment of a baryon
$B$ of this type
\begin{equation}
\mu_B =\frac{4}{3}(\mu_1 + \Delta\mu_1^L) -
      \frac{1}{3}(\mu_3 + \Delta\mu_3^L) .
\end{equation}
$\Delta\mu_1^L$ and $\Delta\mu_3^L$ are the contributions from
$L \neq 0$ configurations in the baryon wave function (Eq.\ (\ref{completewf})),
\begin{eqnarray}
\frac{\Delta\mu_3^L}{\mu_3} &=& \frac {m_3}{M}\ b_0^2  + c_0^2 +
\frac{m_3 -4m_1}{M}\ d_0^2 -3e_0^2 - \frac{m_1}{M}b_0e_0{\cal R} \ , \nonumber \\
 \frac{\Delta\mu_1^L}{\mu_1}&=& \frac{1}{4}( \frac{\Delta\mu_3^L}{\mu_3} +
b_0^2 + 3e_0^2) ,
\end{eqnarray}
with ${\cal R}$ an overlap integral of order unity. The results are quadratic in
the small coefficients $b_0,\ldots,e_0$ in Table \ref{table:wfcoeff} and are
negligible on the scale of the other contributions.


\subsection{\label{subsec:loops}Meson loop corrections}

As noted earlier, the meson loop corrections to the baryon moments have been considered
by many authors \cite{M1,RSS,F,RGG,Caldi,GSS,M2,K,J,L,M3,D,Mei,DH-ChPTmoments,%
Durand,DH-loop-moments,M4,DHJ1,DHJ2} in the context of chiral perturbation theory. We
will use the loop corrections obtained in \cite{DH-loop-moments}. In that paper, the
one-loop corrections were calculated using a meson-nucleon form factor to control the
high-energy behavior of the loop integrals, which is not reliable in the context of
a strict expansion in the chiral parameter $m_s$. The results we obtained after
correcting an error  in \cite{DH-loop-moments} are given in
Table \ref{table:mu_parameters} as decomposed in terms of the parameters
$\mu_a,\ldots,\mu_g$ defined in Eq.\ (\ref{quark-level_moments}). Their connection
with the baryon moments is given in Table \ref{table:B-mu_r}. More detail on the
calculations is given in \cite{DH-loop-moments}.

The meson-loop corrections are distributed over all the parameters $\mu_a-\mu_g$,
corresponding to contributions from the one-, two-, and three- body operators. The
largest parameters, $\mu_c=-0.339$ and $\mu_f=0.471$, arise from the two-body
Thomas-type contributions in Figs.\ \ref{fig1}\,c and \ref{fig1}\,f , the second
a strange-mass correction to the first. There is also a significant three-body
contribution to the moments through $\mu_g$, Fig.\ \ref{fig2}\,g.


\subsection{\label{subsec:analysis}Analysis}

The corrections to the moment parameters for the baryons discussed here are
summarized in Table \ref{table:mu_parameters}. The final rows in the table
give the total calculated corrections to $\mu_a,\ldots,\mu_g,\,\mu_{MM}$,
the values of those parameters obtained by fitting the eight measured moments
exactly, and their difference. The correspondence of the the individual baryon
moments and the $\Sigma^0\Lambda$ transition moment to the parameters is given
in Table~\ref{table:B-mu_r}. The differences for $\mu_a$ and the combination
$\mu_a+\mu_b$ give the values of $3\mu_u/2$ and $-3\mu_s$, the input quark
moments given $\mu_d=-\mu_u/2$.

As was noted in \cite{DH-loop-moments}, fits to the octet moments
using $\mu_u$ and $\mu_s$, or equivalently, $\mu_a$ and $\mu_b$ as
adjustable parameters give only slight improvements relative to
the additive quark model when only the binding corrections, or
only the meson loop corrections are included. Either set of
corrections alone has an incorrect pattern of signs or magnitudes
relative  to the fitted parameters. However, when combined, there
are significant cancellations, and the total result is much closer
to experiment. This may be seen from Table~\ref{table:fit} where
we give the best-fit results for the seven accurately known
moments obtained using the calculated values of $\Delta\mu_{\rm
total}$, and allowing $\mu_a$ and $\mu_b$ to vary. The fit is
quite good, with a mean deviation from experiment of 0.04 $\mu_N$ for
$\mu_p,\ldots, \mu_\Lambda$. The poorly known $\Sigma^0\Lambda$
transition moment has very little weight in a complete fit, and is
left as a prediction,  $\mu_{\Sigma^0\Lambda}=1.522$ $\mu_N$ compared
to the measured value $1.61\pm 0.08$ $\mu_N$.

%
\begin{table*}
\caption{The fit to the baryon magnetic moments obtained using the calculated dynamical
contributions to the chiral parameters $\mu_a,\ldots,\,\mu_{MM}$ given in
Table \protect\ref{table:mu_parameters}, with $\mu_u$ and $\mu_s$ as free parameters.}
\begin{ruledtabular}
\begin{tabular}{ccccccccc}
Baryon & $p$ & $n$ & $\Sigma^+$ & $\Sigma^-$ & $\Xi^0$ & $\Xi^-$ & $\Lambda$ &
$\Sigma^0\Lambda$ \\
\hline
Calculated & 2.744 & -1.955 & 2.461  & -1.069 & -1.278  & -0.598 & -0.607 & $\pm$1.522 \\
Data & 2.793 & -1.913 & 2.458 & -1.160 & -1.250 & -0.651 & -0.613 & $\pm$1.610 \\
 & & & $\pm 0.010$ & $\pm 0.025$ & $\pm 0.014$ & $\pm 0.003$ & $\pm 0.004$ & $\pm 0.080$
\end{tabular}
\label{table:fit}
\end{ruledtabular}
\end{table*}


The advantage of the decomposition in Table~\ref{table:mu_parameters} is the
possible hints it provides for improving the theory. In particular, the most
striking deviation of theory from experiment is in the value of the leading
Thomas-type term $\mu_c$ corresponding to Fig.~\ref{fig1}\,(c). This is well
determined experimentally, with $\mu_n=-\frac{2}{3}(\mu_a-\mu_c)$. The QM
ratio $\mu_n/\mu_p\approx -2/3$ is attained only for $\mu_c$ small, a result
which is ``accidental'' in a general parametrization of the moments \cite{M4,M5}.
Here $\mu_c$ is small because of a cancellation between the meson loop
corrections at one loop and the Thomas-type correction $\Delta\mu^T$.  Some
further input will clearly be needed to obtain better agreement between theory
and experiment.

The calculated value of the related coefficient $\mu_e$, is also significantly
different from experiment. The uncertainties in $\mu_d$, $\mu_g$, and $\mu_{MM}$
are dominated by the relatively large uncertainty in the $\Sigma^0\Lambda$
transition moment. The results nevertheless suggest that the three-body
contribution $\mu_g$ may be overestimated, and the calculated two-strange
quark term $\mu_{MM}$ is too small and of the wrong sign. We note that baryon
mass insertions in the meson loop diagrams, so far not calculated, will affect
these contributions \cite{DHJ2}.


\section{\label{sec:conclusions}SUMMARY}

We have presented and analyzed the QCD corrections to the baryon magnetic moments
in terms of the effective field theory description in terms of ``quark'' operators
${\mb}_a,\ldots,{\mb}_{MM}$ to be used in baryon matrix elements of the form
$\bmu_B=(\bar{B}\bmu B)$, with $B_{ijk}^\gamma$ the effective field operator for
the baryon in Eq.~(\ref{Bijk}). We have then used the connection of this quark
picture to dynamical models \cite{M2,M3,DHJ1,DHJ2} to estimate the input values
of the unknown parameters in the chiral expansion using a semirelativistic dynamical
model. This model can be derived in a quenched approximation to QCD \cite{brambilla}
and gives a good description of the baryon spectrum up to $\sim 3$ GeV \cite{kogut,isgur}.
This allows us to estimate the changes in the leading QM moments in different baryons
and binding corrections to the moments identified earlier \cite{Durand}, and to
study the contributions of orbital angular excitations.

We find that the binding corrections to the simple QM picture of the moments
involve, as expected, mainly contributions from the one- and two- body operators,
and are important. The changes in the QM moments in different baryons are significant
but small.  The orbital contributions to the
moments are completely negligible since the terms in the wave functions which
involve higher angular momenta are very small. The chiral meson loop contributions
are large, but tend to cancel
with the corresponding parameters for the binding corrections leading to a fairly
successful overall fit to the moment data. Fits using the loop corrections alone
are much less satisfactory. The deviations of the calculated parameters
${\mb}_a,\ldots,{\mb}_{MM}$ from those found in an exact fit to the baryons
moments suggests where further input will be important.

We note finally that the procedure used in our construction of  the
$j=\frac{1}{2}^+$ baryon wave functions displays their symmetry properties and
the internal angular momentum content very simply, and may be useful in other
contexts as may the trace methods used to calculate the
many spin matrix elements which appear in the expressions for the magnetic
moments.

\begin{acknowledgments}

One of the authors (PH) is grateful to the Department of Physics,
Creighton University, for its hospitality and support of work done there. The
other author (LD) would like to thank the Aspen Center for Physics for its
hospitality while parts of this work were done.
This work was supported in part by the U.S. Department of Energy under
Grant No.\ DE-FG02-95ER40896, and in part by the University of Wisconsin
Graduate School with funds granted by the Wisconsin Alumni Research Foundation.
\end{acknowledgments}

\appendix

\section{\label{app:jacobi}JACOBI COORDINATES AND RELATIONS}

With $i \ne j \ne k$ and $i$, $j$, and $k$ run from 1 to 3,
we define the Jacobi-type coordinate systems as follows:

\subsection{The space coordinates}

\begin{eqnarray}
{\rb}_{ij} &=& \bx_i - \bx_j \, , \ \
{\bf R}_{ij} = \frac{m_i\bx_i + m_j\bx_j}{m_{ij}} \, , \nonumber \\
{\rb}_{ij,k} &=& {\bf R}_{ij} - \bx_k \, = \,
\frac {m_i(\bx_i - \bx_k) + m_j(\bx_j - \bx_k)}{m_{ij}} \, ,
\nonumber \\
{\bf R}_{ijk} &=& \frac {m_{ij}{\bf R}_{ij} + m_k\bx_k}{M} \, .
\end{eqnarray}
Here $m_{ij}=m_i+m_j$, $M=m_i+m_j+m_k$, and ${\bf R}_{ijk}$ is the usual
center-of-mass coordinate. With these definitions, the roles of
$i$, $j$, and $k$ are completely symmetric.

Thus, if one defines coordinates $\brho = {\rb}_{12}$,
$\blam = {\rb}_{12,3}$,
$\brho' = {\rb}_{23}$,
$\blam' = {\rb}_{23,1}$,
$\brho'' = {\rb}_{31}$, and
$\blam'' = {\rb}_{31,2}$
then $\brho'$, $\blam'$ and $\brho''$, $\blam''$ can be expressed in terms of
$\brho$ and $\blam$ and conversely. For example, if $m_1=m_2$, we find that
\begin{eqnarray}
\brho' &=& \blam -\frac{\brho}{2} \, , \ \
\blam' = -\frac{1}{m_{23}} \left(m_3 \blam + \frac{M}{2} \brho\right) \, ,\\
\brho'' &=& - \blam -\frac{\brho}{2} \, , \ \
\blam'' = -\frac{1}{m_{31}} \left(m_3 \blam - \frac{M}{2} \brho\right) \, .
\end{eqnarray}
One can therefore work with any of the pairs $\brho$,  $\blam$ or
$\brho'$, $\blam'$ or $\brho''$, $\blam''$, and switch between them as necessary.
The spatial volume element is simply $d^3R\,d^3\rho\,d^3\lambda$, or equivalently
for the other pairs of internal coordinates.


\subsection{The momentum coordinates}

\begin{eqnarray}
{\pb}_{ij} &=& \frac{m_j{\pb}_i - m_i{\pb}_j}{m_{ij}}  \, , \ \ \
{\bf P}_{ij} = {\pb}_i + {\pb}_j  \, , \nonumber \\
{\pb}_{ij,k} &=& \frac{m_k{\bf P}_{ij} - m_{ij}{\pb}_k}{M} \, , \ \
{\bf P}_{ijk} = {\bf P}_{ij} + {\pb}_k  \, ,
\end{eqnarray}
where ${\bf P}_{ijk}\equiv{\bf P}$ is the total momentum.

Now, if one denotes
${\pb}_\rho = {\pb}_{12}$, ${\pb}_\lambda = {\pb}_{12,3}$,
${\pb}_{\rho '} = {\pb}_{23}$, ${\pb}_{\lambda '} = {\pb}_{23,1}$,
${\pb}_{\rho ''} = {\pb}_{31}$, and
${\pb}_{\lambda ''} = {\pb}_{31,2}$ then one can choose to work
with either ${\pb}_\rho $, ${\pb}_\lambda$ or
${\pb}_{\rho '}$, ${\pb}_{\lambda '}$ or
${\pb}_{\rho ''}$, ${\pb}_{\lambda ''}$ since there exist relations
between them. For example,
when $m_1=m_2=m$ the coordinates
${\pb}_{\rho '}$, ${\pb}_{\lambda '}$ and
${\pb}_{\rho ''}$, ${\pb}_{\lambda ''}$ can be expresses in terms of
${\pb}_\rho $, ${\pb}_\lambda$ as
\begin{eqnarray}
\frac{{\pb}_{\rho '}}{m_{\rho '}} &=& \frac{{\pb}_\lambda}{m_\lambda}
- \frac{{\pb}_\rho}{2m_\rho} \, , \ \
{\pb}_{\lambda '} = -{\pb}_\rho - \frac{{\pb}_\lambda}{2} \, ,
\nonumber \\
\frac{{\pb}_{\rho ''}}{m_{\rho ''}} &=& -\frac{{\pb}_\lambda}{m_\lambda}
- \frac{{\pb}_\rho}{2m_\rho} \, , \ \
{\pb}_{\lambda ''} = {\pb}_\rho - \frac{{\pb}_\lambda}{2} \, ,
\end{eqnarray}
where $m_\rho =m_1m_2/m_{12}=2/m$, $m_\lambda=m_3 m_{12}/m_{123}$,
$m_{\rho '}=m_2m_3/m_{23}$, and $m_{\rho ''}=m_3m_1/m_{31}$. The volume element
in momentum space is $d^3P\,d^3p_\rho\,d^3p_\lambda$, and equivalently for the
other pairs of internal momenta.


\section{\label{app:traces}TRACE RELATIONS AND INDEPENDENT OPERATORS}

We present here some trace relations derived from the trace method given in
subsection \ref{subsec:traces}. These results are useful to an evaluation
of the spin matrix elements. We also use them to prove the equivalence of some
states shown in Eq.(\ref{allstates}).


\subsection{\label{app:spin_traces}Results for spin traces}

Let $\bA$, $\bB$, and $\bC$ be any spin-independent vectors and
$\bx$, $\by$, and $\bz$ be the coordinate-type vectors (such as
$\brho$, $\brho'$, $\blam$, ...), we have
\begin{eqnarray}
{\rm Tr} \, (\bsig_1-\bsig_2) \cdot
\bsig_3 \, \bsig_1 \cdot \bsig_2
\, P_{\frac{1}{2},1} & = & 0  \, ,\\
{\rm Tr} \, (\bsig_1-\bsig_2) \cdot
\bsig_3 \, \bsig_2 \cdot \bsig_3
\, P_{\frac{1}{2},1} & = & -12 \, , \\
{\rm Tr} \, (\bsig_1-\bsig_2) \cdot
\bA \, \bsig_i \cdot \bsig_j
\, P_{\frac{1}{2},1} & = & 0 \, , \\
{\rm Tr} \, (\bsig_1-\bsig_2) \cdot
\bsig_3 \, \bsig_i \cdot \bA
\, P_{\frac{1}{2},1} & = & 0 \, , \\
{\rm Tr} \, (\bsig_1-\bsig_2) \cdot
\bA\, (\bsig_1-\bsig_2) \cdot \bB
\, P_{\frac{1}{2},1} & = & \frac{8}{3} \bA \cdot \bB  \, ,\\
{\rm Tr} \, (\bsig_1-\bsig_2) \cdot
\bA\, (\bsig_1-\bsig_2) \cdot \bB\, \bsig_3 \cdot \bC
\, P_{\frac{1}{2},1} & = & - \frac{8i}{3} (\bA \times \bB) \cdot \bC \, , \\
{\rm Tr} \, (\bsig_1-\bsig_2) \cdot \bA\, \bsig_j \cdot \bB \,
(\bsig_1-\bsig_2) \cdot
\bA\, P_{\frac{1}{2},1} & = & 0 \, \, , (j=1,2,3) \, , \\
{\rm Tr} \, t_{ij}(\bx) \bsig_k \cdot \bA \, P_{\frac{1}{2},1}
& = & 0  \, \, ,  (i, j, k = 1, 2, 3) \, ,\\
{\rm Tr} \, t_{12}^2(\bx) \, P_{\frac{1}{2},1} & = & 16 \bx^4 \, , \\
{\rm Tr} \, t_{23}^2(\bx) \, P_{\frac{1}{2},1} & = & 4 \bx^4 \, , \\
{\rm Tr} \, t_{12}(\bx) t_{23}(\by)
\, P_{\frac{1}{2},1}  =  {\rm Tr} \, t_{23}(\by)
t_{12}(\bx)\, P_{\frac{1}{2},1} & = &
4 [\bx^2 \by^2 - 3 (\bx \cdot \by)^2] \, , \\
{\rm Tr} \, t_{23}(\bx) t_{31}(\by)
\, P_{\frac{1}{2},1}  =  {\rm Tr} \, t_{31}(\by)
t_{23}(\bx)\, P_{\frac{1}{2},1} & = &
- 4 [\bx^2 \by^2 - 3 (\bx \cdot \by)^2] \, .
\end{eqnarray}


\subsection{\label{app:redundant}Proof of the equivalence of redundant states}

First consider states $|\psi_1 \rangle$ and $|\psi_2 \rangle$ defined in terms
of the two operators in Eq.\ (\ref{row3}) by
\begin{eqnarray}
| \psi_1 \rangle &=& |(\bsig_1-\bsig_2) \times
\bsig_3 \cdot \brho \times
\blam \,\chi_{j_3}^{S_{12}=1} \rangle \nonumber \\
&=& |[(\bsig_1-\bsig_2)\cdot
\brho\, \bsig_3 \cdot \blam
-(\bsig_1-\bsig_2)\cdot
\blam\, \bsig_3 \cdot \brho]\,
\chi_{j_3}^{S_{12}=1} \rangle  \, , \\
| \psi_2 \rangle & = & | i(\bsig_1-\bsig_2) \cdot
\brho \times \blam\,\chi_{j_3}^{S_{12}=1} \rangle \, ,
\end{eqnarray}
where the
spatial wave functions are suppressed for simplicity. Using the trace identities
shown above, it is easy to show that for $j=\frac{1}{2}$
\begin{equation}
\langle \psi_1 | \psi_1 \rangle = 4 \langle \psi_1 | \psi_1 \rangle
= 2 \langle \psi_1 | \psi_2 \rangle = 2 \langle \psi_2 | \psi_1 \rangle
= \frac{16}{3}(\brho \times \blam)^2 \, .
\end{equation}
Thus, if we choose a state $|\psi \rangle$ such that $|\psi \rangle =
|\psi_1 - 2 \psi_2 \rangle $,
then $\langle \psi | \psi \rangle = 0$ and the states $|\psi_1 \rangle$
and $2|\psi_2 \rangle$ are equivalent. We will therefore drop the second operator
in Eq.\ (\ref{row3}) in writing the general baryon wave function.

Next, we show that the states given by the action of the operators $t_{12}(\bx)$ and
$t_{23}(\bx)+t_{31}(\bx)$ in Eq.\ (\ref{row4}) on $\chi_{j_3}^{S_{12}=1}$ are
also not independent for $j=\frac{1}{2}$. Let us consider the following state
\begin{equation}
| \psi \rangle = | (t_{12}(\bx) + t_{23}(\bx)+
 t_{31}(\bx))\chi_{j_3}^{S_{12}=1} \rangle \, .
\end{equation}
A calculation of $\langle \psi | \psi \rangle$ using the trace method yields
\begin{eqnarray}
\langle \psi | \psi \rangle &=& \frac{1}{2} {\rm Tr}
\,  [t_{12}(\bx) + t_{23}(\bx)+
 t_{31}(\bx)]^2 P_{\frac{1}{2},1} \nonumber \\
 &=&\frac{1}{2} {\rm Tr}
\, [t_{12}(\bx) + 2 t_{23}(\bx)]^2 P_{\frac{1}{2},1}
\nonumber \\
 &=&\frac{1}{2} {\rm Tr}
\,  [t_{12}^2(\bx) + 4 t_{23}^2(\bx)
+ 2t_{12}(\bx)t_{23}(\bx)+
2t_{23}(\bx)t_{12}(\bx)]
 P_{\frac{1}{2},1} \nonumber \\
&=& \frac{1}{2}[ 16 \bx^2+16 \bx^2
+ 2(-8 \bx^2) + 2(-8 \bx^2)] = 0 \, .
\end{eqnarray}
Hence, the spin functions $t_{12}(\bx)\chi_{j_3}^{S_{12}=1}$ and
$[t_{23}(\bx)+t_{31}(\bx)]\chi_{j_3}^{S_{12}=1}$ in Eq.\ (\ref{row4}) are
equivalent. The same result holds for the operators in Eq.\ (\ref{row5}).


\section{\label{app:matrix_elements}CALCULATION OF MATRIX ELEMENTS}

Let us consider the spatial matrix element
$ < \psi_0^A| V | \psi_0^B)>$, $A$,$B = a$, $b$, $c$, $d$, and
$e$. Here $V=V_0(r_{ij})P(\brho,  \blam)$
with $V_0$ is a function of the absolute
value of ${\rb}_{ij}$ and $P(\brho,  \blam)$
is a polynomial of $\brho$ and $\blam$. The Gaussian
wave functions $\psi (\beta_A)$
are defined by Eq.\ (\ref{gausswf}). Then, one has
\begin{eqnarray} \label{exp1}
\langle \psi_0^A| V | \psi_0^B\rangle &=&
( \frac{\beta_\rho^A \beta_\lambda^A \beta_\rho^B \beta_\lambda^B}{\pi^2} )^{3/2} \\
&& \times
\int d \brho \, d \blam
V_0(r_{ij})P(\brho, \blam)
 \exp{ [ -(\tilde{\beta}_\rho^2 \rho^2 +
\tilde{\beta}_\lambda^2 \lambda^2)]} \, ,
\nonumber
\end{eqnarray}
where
\begin{equation}
\tilde{\beta}_\rho^2 = \frac{1}{2}({\beta_\rho^A}^2 + {\beta_\rho^B}^2) \, ,
\ \  \tilde{\beta}_\lambda^2 =
\frac{1}{2}({\beta_\lambda^A}^2 + {\beta_\lambda^B}^2) \, .
\end{equation}
If $r_{ij}=r_{12}=\rho$, then the integration is straightforward.

Now consider the case when
\begin{equation}
{\rb}_{ij} = a \blam + b \brho \, ,
\end{equation}
where $a$, $b$ are constants. If one denotes ${\rb}_{ij} = {\sbd}$, then
$\blam= ({\sbd} - b \brho)/a $ and
\begin{equation}
\tilde{\beta}_\rho^2 \rho^2 + \tilde{\beta}_\lambda^2 \lambda^2 =
\tilde{\beta}_\rho^2 \left(1+\tilde x \frac{b^2}{a^2}\right)(\brho -
w {\sbd})^2 + \frac{\tilde{\beta}_\lambda^2}{a^2 + \tilde x b^2} {\sbd}^2
\, ,
\end{equation}
where $\tilde x = {\textstyle{\tilde{\beta}_\lambda^2}/
\textstyle{\tilde{\beta}_\rho^2}}$ and
$w = {\textstyle{\tilde x b}/(\textstyle{a^2 + \tilde x b^2}})$ \, .
A translation $\brho \rightarrow \brho + w {\sbd}$
gives the result $\blam=(\tilde w {\sbd}+b \brho)/a$
with $\tilde w = 1-bw$. Eq.\ (\ref{exp1}) now becomes
\begin{eqnarray} \label{exp2}
\langle \psi_0^A| V | \psi_0^B \rangle &=&
\left( \frac{\beta_\rho^A \beta_\lambda^A \beta_\rho^B \beta_\lambda^B}{\pi^2}
\right)^{3/2} \\
&& \times
\int \frac{d \brho \, d {\sbd}}{|a|^3}
V_0(|{\sbd}|)P\left(\brho+w{\sbd}, (\tilde{w}{\sbd}+b\brho)/a\right)
 \exp{ [ -(\beta_\rho^{'2} \rho^2 +
\beta_\lambda^{'2} {\sbd}^2)]} \, ,
\nonumber
\end{eqnarray}
where
\begin{equation}
\beta_\rho^{\,'2} = {\tilde\beta_\rho}^2\left(1 + \tilde x \frac{b^2}{a^2}\right) \, ,
\ \  \beta_\lambda^{\,'2} =
\frac{\tilde\beta_\lambda^2}{a^2 + \tilde x b^2} \, .
\end{equation}
At this step, the integration of the right hand side of Eq.\ (\ref{exp2})
becomes straightforward.

For illustration, let us consider a simple case when
$\psi_0^A = \psi_0^B = \psi_0^a$ and $V=(\brho \times \blam)^2/ \rho '$. In this
case, $V_0(r_{ij})=1/\rho '$ and $P(\brho,  \blam)=(\brho \times \blam)^2$.
Since $\brho' = \blam - \brho/2$, one has $a=1$, $b=-1/2$. Also in this
case $\tilde\beta_\rho = \beta_\rho$, $\tilde\beta_\lambda = \beta_\lambda$,
$\tilde x = x = {\beta_\lambda^a}^2 / {\beta_\rho^a}^2$, and $ w = -2x/(4+x)$.
Note that under the translation $\brho \rightarrow \brho + w {\sbd}$,
the expression $(\brho \times \blam)^2$ transforms into $(\brho \times {\sbd})^2$
that can be replaced by $\frac{2}{3}\rho^2 s^2$ after working out the
angular dependent parts. One finds
\begin{eqnarray} \label{exp3}
\langle \psi_0^a| \frac{(\brho \times \blam)^2}{\rho '} | \psi_0^a \rangle &=&
\left( \frac{\beta_\rho^a \beta_\lambda^a}{\pi} \right)^3
\int d \brho \, d {\sbd}\, \frac{(\brho \times {\sbd})^2}{s} \,
 \exp{ [ -(\beta_\rho^{\,'2} \rho^2 +
\beta_\lambda^{\,'2} {\sbd}^2)]} \, \nonumber \\
&=& \frac{2}{\sqrt{\pi}}\,\frac{1}{\beta^{\,'2}_\rho\beta^{\,'}_\lambda} \, ,
\end{eqnarray}
where
\begin{equation}
\beta_\rho^{\,'2} = {\beta_\rho^a}^2\left(1 + \frac{x}{4}\right)  ,
\qquad \beta_\lambda^{\,'2} =
{\beta_\lambda^a}^2 / \left(1 + \frac{x}{4}\right) ,
\qquad \beta^{\,'}_\lambda\beta^{\,'}_\rho=\beta_\lambda\beta_\rho.
\end{equation}

We can use similar methods with the momentum-space wave functions
\begin{equation}
\label{momentum_wf}
\tilde{\psi}_0^a({\pb}_\rho,{\pb}_\lambda) =
\left(\frac{1}{\pi\beta_\rho\beta_\lambda}\right)^{3/2}
\exp{\left(-\frac{p_\rho^2}{2\beta_\rho^2}-\frac{p_\lambda^2}{2\beta_\lambda^2} \right)}
\end{equation}
to evaluate integrals which involve the energies $E_i=\sqrt{p_i^2+m_i^2}$. Thus,
\begin{eqnarray}
\Big\langle \sqrt{p_1^2+m_1^2}\Big\rangle &=& \beta^{\,'}_\rho I_0
\left(\frac{m_1^2}{\beta^{\,'2}_\rho}\right),\qquad \Big\langle
\sqrt{p_3^2+m_3^2}\Big\rangle =
\beta_\lambda I_0\left(\frac{m_3^2}{\beta_\lambda^2}\right),
\\
\Big\langle 1/\sqrt{p_1^2+m_1^2}\Big\rangle &=&
\frac{1}{\beta^{\,'}_\rho} J_0\left(\frac{m_1^2}{\beta^{\,'2}_\rho}\right),
\qquad \Big\langle 1/\sqrt{p_3^2+m_3^2}\Big\rangle
= \frac{1}{\beta_\lambda} J_0\left(\frac{m_3^2}{\beta_\lambda^2}\right),
\end{eqnarray}
where
\begin{equation}
\label{I1,I2}
I_n\left(z^2\right) = \frac{4}{\sqrt{\pi}}\int_0^\infty dt\;t^{2n+2}\sqrt{t^2+z^2}
e^{-t^2},\qquad J_n\left(z^2\right) =
\frac{4}{\sqrt{\pi}}\int_0^\infty dt\,\frac{t^{2n+2}}{\sqrt{t^2+z^2}} e^{-t^2}.
\end{equation}
The integrals $I_n$ and $J_n$ must be evaluated numerically.

Working out the spin matrix elements and applying the technique of integration
shown above, one can evaluate the matrix elements
$ \langle \psi_{\frac{1}{2},m}| H | \psi _{\frac{1}{2},m} \rangle$ .
For example, the matrix element $ \langle \psi_0^a| H_0 | \psi_0^a \rangle$
is given by
\begin{eqnarray}
\Big\langle \psi_0^a| H_0 | \psi_0^a \Big \rangle &=&
2\beta^{\,'}_\rho I_0\left(m_1^2/\beta^{\,'2}_\rho\right) +
\beta_\lambda I_0\left(m_3^2/\beta_\lambda^2\right)
\\
&&+ \frac{\sigma}{\sqrt{\pi}}\left( \frac{1}{\beta_\rho}
+\frac{2}{\beta^{\,'}_\lambda}\right)  -
\frac{4\alpha_{\rm s}}{3} \frac{1}{\sqrt{\pi}}\left( \beta_\rho
+ 2 \beta^{\,'}_\lambda\right)
\, .
\nonumber
\end{eqnarray}
The matrix elements of $H_0$ in the states in Eq.~(\ref{wavefunc}) with
$L_\rho,\;L_\lambda\not=0$ involve the $I_n$ with $n=0,\,1,\,2$.

\end{document}